\documentclass[submission,copyright,creativecommons]{eptcs}
\providecommand{\event}{MARS 2022} 
\usepackage{breakurl}             
\usepackage{underscore}           

\usepackage{graphicx}
\usepackage{xcolor} 

\usepackage{soul}

\usepackage[section]{placeins}

\graphicspath{{Images/}}

\title{Formal Modeling and Initial Analysis\\ of the 4SECURail Case Study}
\author{
Franco Mazzanti
\institute{ISTI-CNR\\ Pisa, Italy}
\email{franco.mazzanti@isti.cnr.it}
\and
Dimitri Belli
\institute{ISTI-CNR\\ Pisa, Italy}
\email{dimitri.belli@isti.cnr.it}
}
\def\titlerunning{Formal Modelling and Analysis of the 4SECURail Case Study}
\def\authorrunning{F. Mazzanti \& D. Belli}
\begin{document}
\maketitle

\begin{abstract}
We present the case study developed in the context of the 4SECURail  project and the approach used for its formal modeling and analysis.
Starting from a simple SysML/UML behavioral model of the system requirements, three formal models have been developed using three different frameworks, namely UMC, ProB, and CADP/LNT. The paper shows how the different ways to represent and analyze the system from the three different points of view allow us to take advantage of the resulting diversity.
\end{abstract}

\section{Introduction}


One of the goals of the 4SECURail project\footnote{https://4SECURail.eu  November 2019 -- November 2021.} is to observe the possible approaches, benefits, limits, and costs of introducing formal methods inside the \emph{requirements definition} process in the context of railway-signaling systems.
This has been done with the set up of a ``demonstrator'' with the purpose to exemplify the application of state-of-the-art tools and methodologies to a selected railway case study with the collection of meaningful information on the costs and benefits of the process. 
The overall context and objectives of this project and experimentation are described in \cite{refD2.1,refrssrail}; in this paper, we describe specifically the approach that has been followed for the formal modeling and initial analysis of the case study, which has seen the exploitation of three different formal verification frameworks.
The rest of the paper is structured as follows: In Section 2, we provide details about the case study that has been the object of the experimentation; in Section 3, we present the formal modeling approach that has been adopted in the demonstrator process; in Section 4, we describe the various kinds of analysis performed. In Sections 5 and 6, we respectively hint at some related works and draw our conclusions.


\section{The reference case study}

\begin{figure}[ht!]
\centering
\includegraphics[width=0.75 \textwidth]{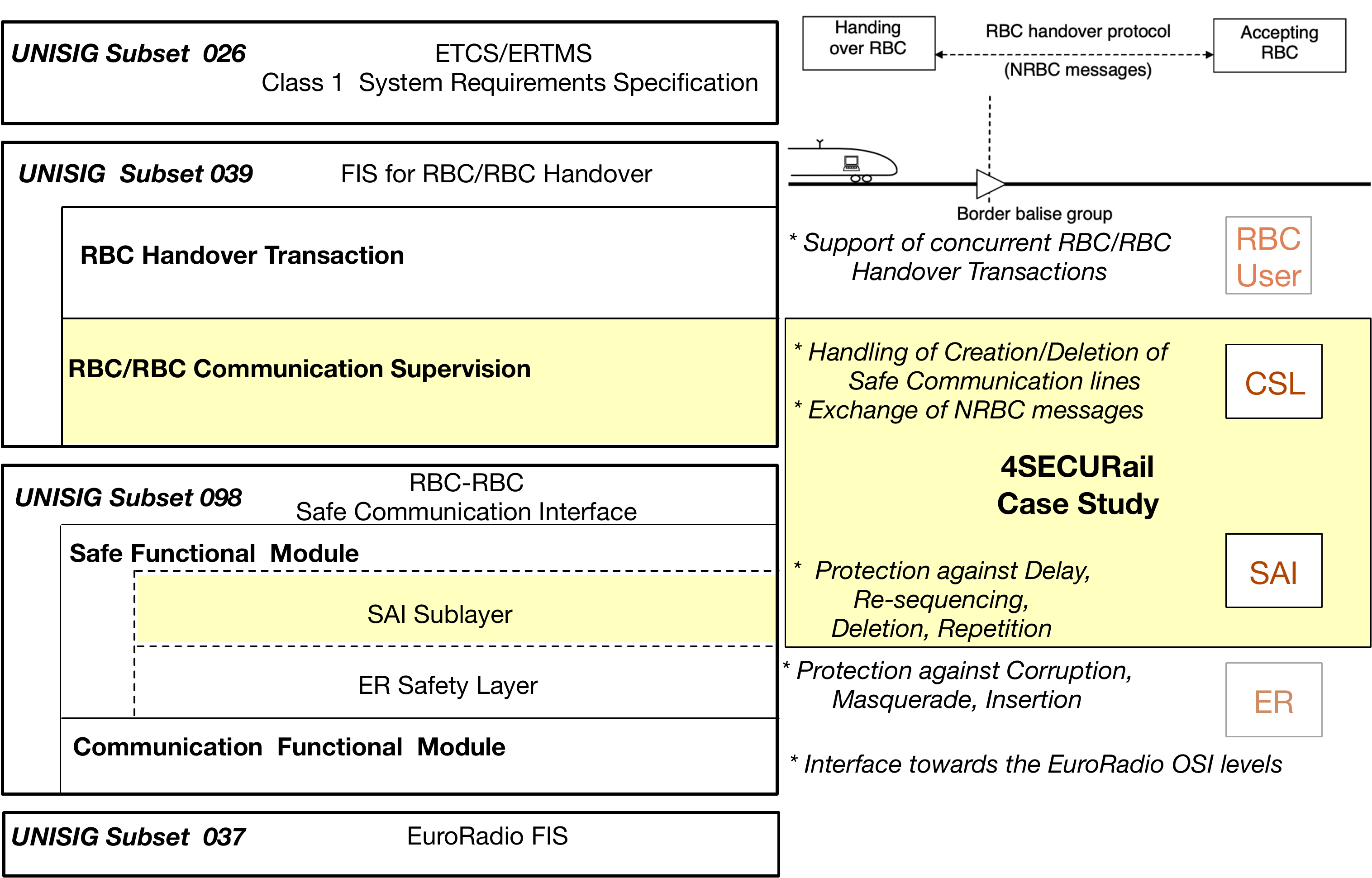}
\caption{Overall structure of the 4SECURail case study} \label{CASESTUDY}
\end{figure}

The transit of a train from an area supervised by a Radio Block Centre (RBC) to an adjacent area supervised by another RBC occurs during the so-called RBC-RBC handover phase and requires the exchange of information between the two RBCs according to a specific protocol. This exchange of information is supported by the communication layer specified within the UNISIG SUBSET-039~\cite{sub39}, UNISIG SUBSET-098~\cite{sub98}, and UNISIG SUBSET-037~\cite{sub37}, and the whole stack is implemented by both sides of the communication channel. 
Figure~\ref{CASESTUDY} summarizes the overall structure of the UNISIG standards, supporting the handover of a train.
The 4SECURail case study is based on two main sub-components of the communication layers constituting the RBC-RBC handover. The considered components are the Communication Supervision Layer (CSL) of the SUBSET-039 and the Safe Application Intermediate SubLayer (SAI) of the SUBSET-098. These two components are the main actors that support the creation/deletion of safe communication lines and protect the transmission of messages exchanged on such lines.
In particular, the CSL is responsible for requesting the activation -- and in case of failure, the re-establishment -- of the communication line, for continuously controlling its liveliness, and for the forwarding of the handover transaction messages. The SAI is responsible for ensuring the absence of excessive delays, repetitions, losses, or re-ordering of messages during their transmissions. This is achieved by adding sequence numbers and time-related information to the RBC messages.
The RBC/RBC communication line consists of two sides that are properly configured as ``initiator'' and ``called''.
With respect to the SUBSET-098, the 4SECURail case study neither includes the EuroRadio Safety Layer (ER), which is responsible for preventing corruption, masquerading and insertion issues during the communications, nor the lower Communication Functional Module (CFM) interface. With respect to the SUBSET-039, the 4SECURail case study does not include the description of the activation of multiple, concurrent RBC-RBC handover transactions when trains move from a zone supervised by an RBC to an adjacent zone supervised by another RBC.
From the point of view of the CSL, the RBC messages are forwarded to/from the other RBC side without the knowledge of their specific contents or session to which they belong.
The case study of the project, as derived from the above-mentioned standards, is described in natural language in Deliverable D2.3~\cite{refD2.3}, along with the rationale for its choice.
 Of course, the level of abstraction of these requirement documents is not that one of an executable system specification, but a higher level. 

\section{The formal modeling}
\subsection{From natural language to executable UML specifications}

\begin{figure}[ht!]
\centering
\includegraphics[width=0.7 \textwidth]{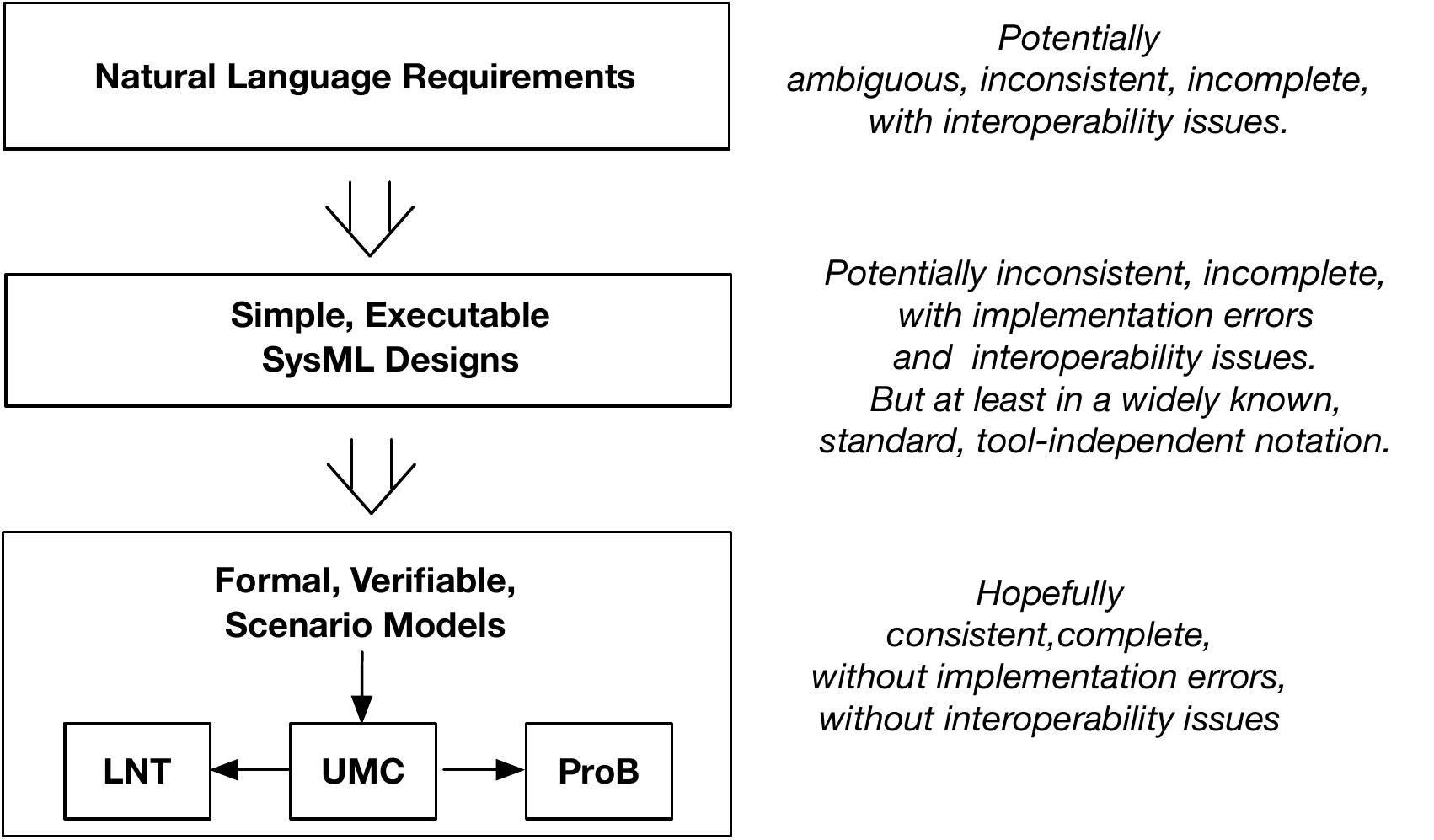}
\caption{From natural language to formal models} \label{FromTo}
\end{figure}

As shown in Figure~\ref{FromTo}, the first step towards the generation of formal models of the system is the description -- in terms of extremely simple SysML/UML features -- of the system components described by the natural language requirements. It is well known that requirements described in free-style natural language suffer the risk of being unclear (e.g., redundant), potentially ambiguous, in part contradictory, and possibly not describing essential aspects. Moreover, since the railway infrastructure is essentially a system of systems, specifying and guaranteeing the desired interoperability among the various components is a more challanging task than specifying and guaranteeing the independent safety of each singularly specified component.

Constructing a possible implementation using a potentially executable notation \emph{with clear semantics} allows (at least) to remove the underlying ambiguities but, until the system is thoroughly tested or verified, the risk of logical deficiencies persists. However, beyond the beneficial ``natural language interpretation'' step, the ``executable implementation'' step risks being a critical source of mistakes. 
Therefore, the subsequent formal modeling and analysis step of the ``executable implementation'' becomes essential.
The association of the term ``clear semantics'' with the term ``UML'' can be, in general, quite problematic. 
In our case, we have used the very minimal set of UML features needed for our executable modeling, avoiding all the complexities related, for example, to composite states, transition priorities, deferred events, and making the explicit assumption of FIFO event queues. The extreme simplicity of the resulting subset aims not only to the association of a ``precise semantics'' and a \emph{simple intituive meaning} to the designs but also to an ``easy translation'' of the designs into several formal notations.
Appendix A shows our reference UML state-machine diagrams for the CSL and SAI system components of the case study, in both their \emph{initiator-side} and \emph{called-side} version.

\subsection{From executable UML specifications to verifiable scenarios}

\begin{figure}[ht!]
\centering
\includegraphics[width=0.4 \textwidth]{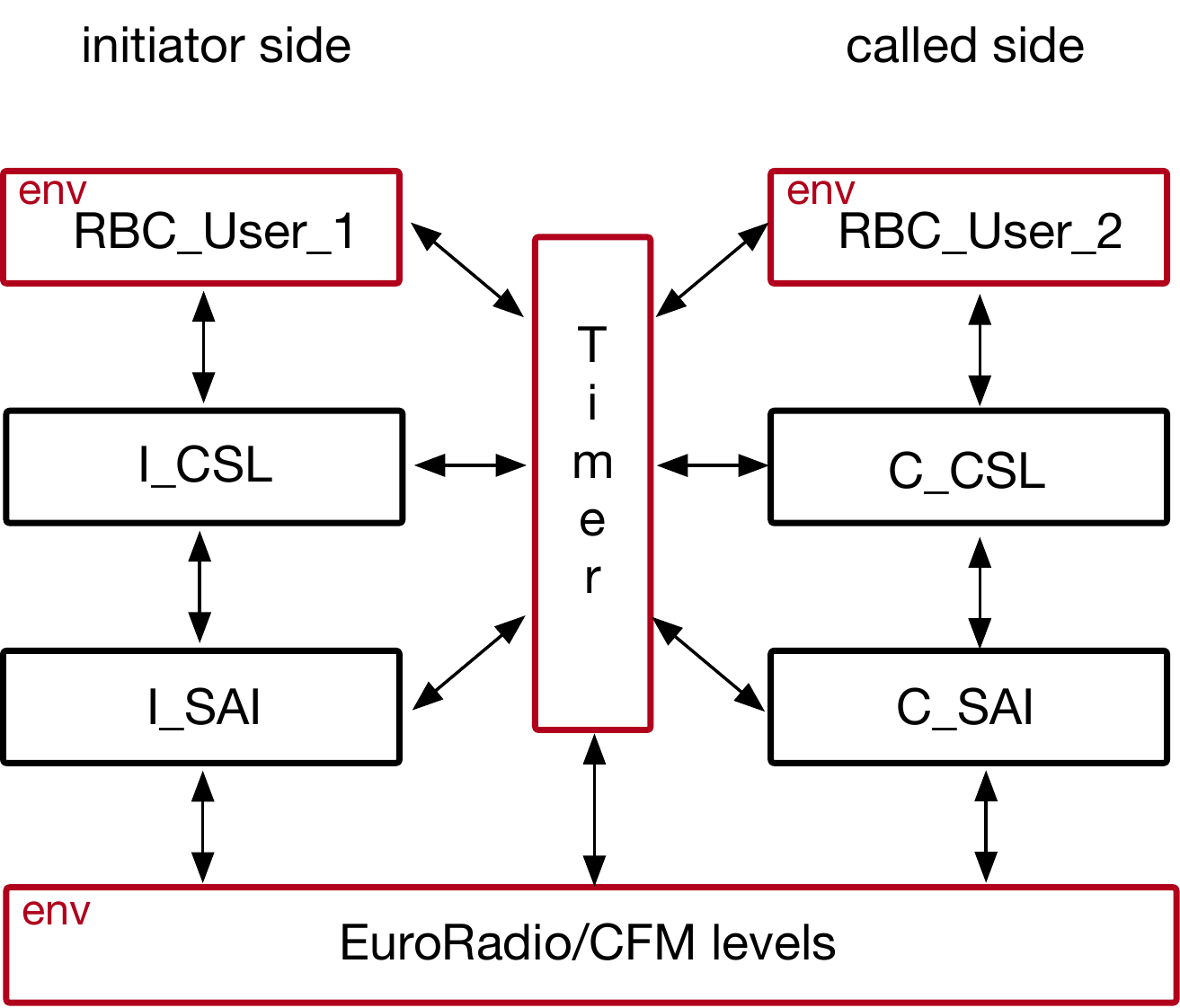}
\caption{The complete executable system structure}  \label{systemcomponents}
\end{figure}

The system requirements in the Deliverable D2.3~\cite{refD2.3} have been the base for the design of the executable models of the CSL and SAI components. However, in order to have an actually verifiable system, we need a \emph{closed} system that contains the specified components plus all the needed environment components that stimulate, receive data, and forward messages from the initiator to called side of the system.  In order to deal with the time-related aspects of the specification, we also introduce a timer component that allows all the other components to proceed in parallel in an asynchronous way but relatively at the same speed\footnote{Since all the system components are modeled as executing a cyclic activity, the timer component just constrains the  frequency of the cycles to be the same while allowing the overlapping of their behavior.}.
Figure ~\ref{systemcomponents} shows the resulting structure of the whole system. Also all the added environment and timer components can be designed in UML to facilitate the system encoding into the selected formal notations. An example of these environment components is contained in Appendix A.

It is infeasible/impractical to define these environment components in their full possible generality because the system components are heavily dependent on several configuration parameters. It makes more sense to define them according to the properties we intend to verify on the complete closed system.
As first step of formal modeling, the executable UML system diagrams corresponding to a given scenario are translated into the notation accepted by the UMC\footnote{https://fmt.isti.cnr.it/umc} tool.  
At the beginning of the project, the possibility of designing the SysML system using a commercial MBSE framework -- namely SPARX-EA\footnote{https://sparxsystems.com/products/ea/index.html} -- has been evaluated. 
This approach has been abandoned because of time and effort constraints of the project. Implementing a translator from the SPARX-generated XMI towards UMC would have been a significant effort. Moreover, it would have tied the whole analysis approach to a specific commercial tool, a fact which was not considered desirable. 
Therefore our initial SysML models have the structure of simple graphical designs; their role is just that one of constituting an intermediate, easy-to-understand documentation halfway between the natural language requirements and the formal models\footnote{more details can be found in \cite{refrssrail}}.
Starting from the UMC notation, further formal models have been automatically generated in the notations accepted by the ProB\footnote{https://prob.hhu.de/} and CADP/LNT\footnote{https://cadp.inria.fr/} tools.
UMC~\cite{refKANDI,refUMC2,refUMC1}  has been chosen as the target of the initial formal encoding because it is a tool natively oriented to fast-prototyping of SysML systems. 
It supports a textual notation of UML state-machine diagrams that directly reflects the graphical counterpart, allows fast state-space exploration, state- and event-based  (on-the-fly) model checking, and detailed debugging of the system. 
Last but not least, it is part of a framework developed locally at ISTI. We have a deep insider knowledge that allowed us to easily implement translators towards the other formal notations within the time and effort constraint of the project. However, UMC is essentially a teaching/research-oriented academic tool and lacks the maturity, stability, and support level required by an industry-usable framework. 
Also for this reason we have planned inside the project the exploitation of further, more industry-ready formal frameworks. 
ProB \cite{refProB2008} has been selected as the second target of the formal encoding because of its recognized role (see e.g. \cite{refICSE,refTSE}) in the field of formal railway-related modeling. Is it supported by (more than one) very user-friendly GUI. It allows LTL/CTL model checking, state-space exploration, state-space projections, and trace descriptions in the form of sequence diagrams. 
Last but not least, it is a framework with which we have already had some previous modeling experience \cite{refD43}, and that did not require a learning-from-scratch step. 
CADP/LNT~\cite{refCADP,refLNT} has been selected as the third target of the formal encoding because of its theoretical roots on LTS-related theories. These allow to reason in terms of minimizations, bisimulations, and compositional verifications. CADP is a rich toolbox that supports a wide set of  $\mu$-calculus-based branching-time logic and a powerful scripting language (SVL~\cite{refSVL}) to support  verification. Also in this case its choice has been influenced by the previous experiences we have had with this framework \cite{refTACAS,refFMSD}.
There are several ways in which SysML/UML designs might be encoded into the ProB and LNT formal notations.
In our case, we made the choice to generate both ProB and LNT models \emph{automatically} from the UMC model. 
The translation implemented in our demonstrator is still a preliminary version and does not exploit at best all the features potentially offered by the target framework \footnote{E.g. all message parameters are mapped into integer values without considering the specific subrange to which that might belong.}. Nevertheless, the availability of the automatic translation proved to be an essential aspect of the demonstrated approach. 
Our models and scenarios have been developed incrementally, with a long sequence of refinements and extensions. At every single step, we have been able to quickly perform the lightweight formal verification of interest with almost no effort. This would not have been possible without an automatic generation of the ProB and LNT models.
In the following, we will give some details on the overall structure of the generated models, referring to D2.5~\cite{refD2.5} for a broader presentation. All the UMC/ProB/LNT models, specifying the scenarios of interest, are available from an open access repository~\cite{refZENmodels}, as well as the source code of the applied translators~\cite{refZENcode}.

\subsubsection{UMC encoding}

\begin{figure}[ht!]
\centering
\includegraphics[width=0.7 \textwidth]{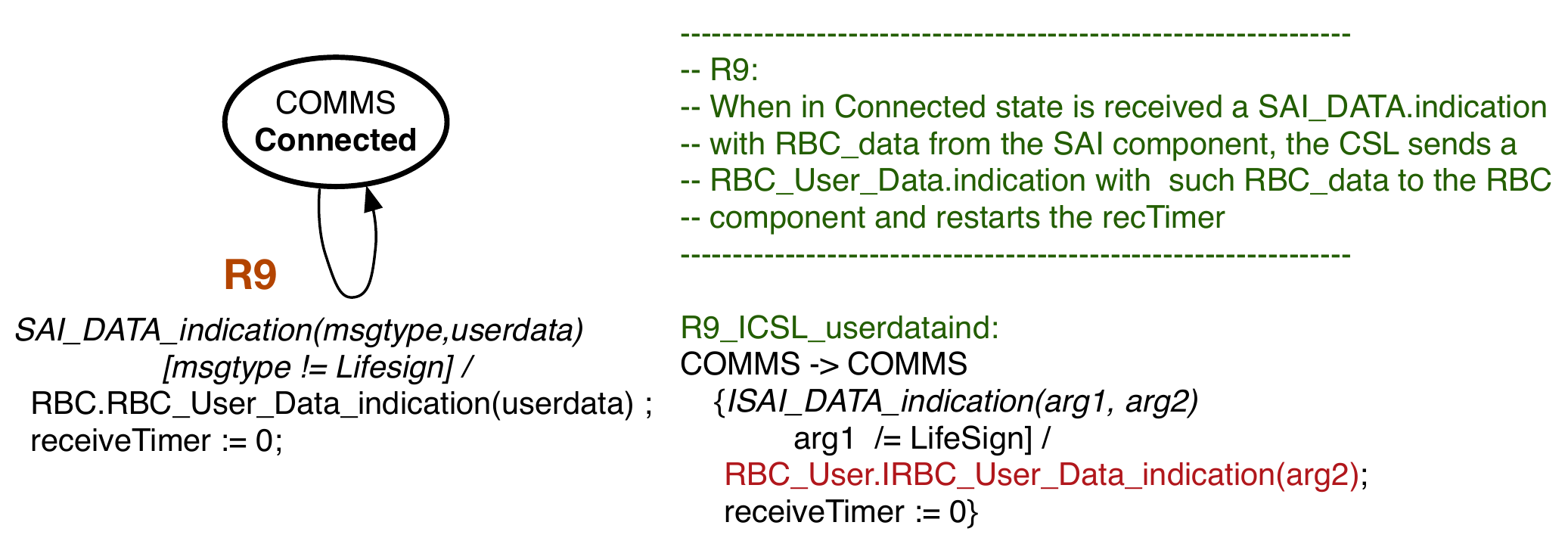}
\caption{Textual encoding of a state-machine transition}  \label{UMCrule}
\end{figure}

In UMC, a system definition is specified as a set of active objects that are instances of class definitions.  A class declaration 
specifies a template of state-machine, defining the set of events accepted by the machine, its local variables, and the state-machine behavior when state transitions are triggered.
State machine transitions are encoded in a simple textual form and specify, as shown in Figure~\ref{UMCrule}:
\begin{itemize}
\vspace{-1pt}
\item  an optional transition label (R9\_ICSL\_userdataind),
\vspace{-3pt}
\item the source and target states of the transition (COMMS, COMMS),
\vspace{-3pt}
\item a block \{...\} containing: the triggering event of the transition (ISAI\_DATA\_indication), possibly with parameters and guards, and the sequence of actions to be performed as an effect of the transition (the sending of the IRBC\_User\_Data\_indication signal to the RBC\_User component and the assignment to the receiveTimer variable).
\end{itemize}

\begin{figure}[ht!]
\centering
\includegraphics[width=0.8 \textwidth]{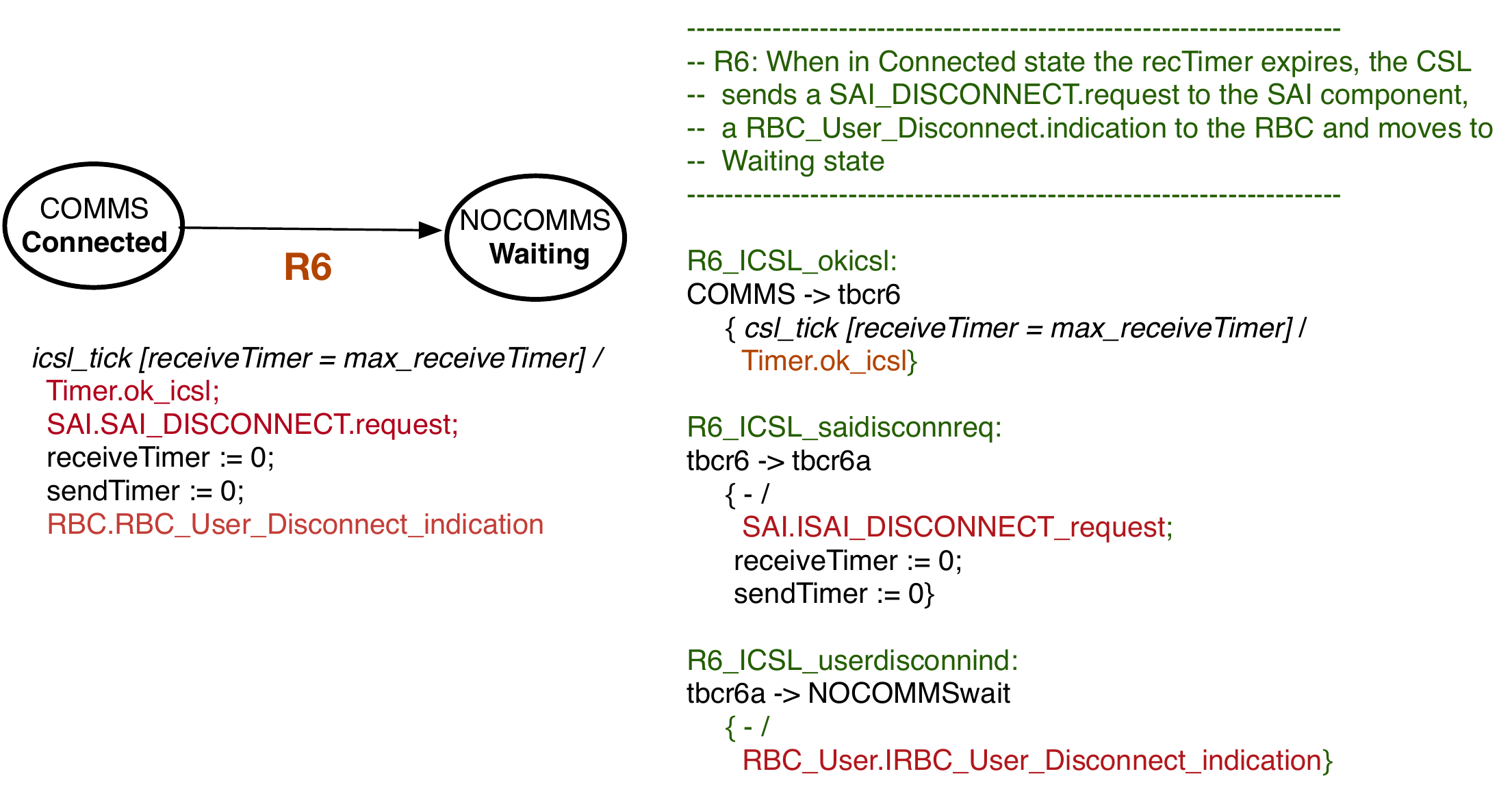}
\caption{Splitting a UML transition into a sequence of UMC atomic transitions}  \label{UMCr6}
\end{figure}

Appendix  B shows the UMC encoding of the component I\_CSL whose UML state-machine diagram is shown in Appendix A.
The mapping of the UML diagrams to the UMC encoding is almost direct. 
There are only a few aspects that deserve some attention. One point is that UMC transitions are ``atomic'' also at the system level, while the UML transitions are ``atomic'' only with respect to the state-machine to which they belong. Therefore, if we have a UML transition that sends several signals to other objects, a correct modeling of the behavior requires splitting the UML transition into several atomic steps. An example of this is shown in Figure~\ref{UMCr6}, where a UML transition sending three signals is split into a sequence of three UMC transitions. The second point is that in UML, when a dispatched event does not trigger any transition it is simply removed from the event queue and discarded. 
This behavior is implicit in the state-machine diagram, but it is reasonable to make it explicit in the UMC designs to simplify the translation of the models into the other notations. This also allows distinguishing more clearly the case in which an event is intentionally (correctly) discarded from the cases in which the arrival of the event is simply a not relevant situation or the case in which it is a really unintended behavior highlighting a case of system malfunctioning.

\vspace{-10pt}

\subsubsection{ProB encoding}

A system specification is structured in ProB as a ``B machine''. In our case, since the system under analysis is composed of several mutually interacting state-machines (and the B language is not able to deal with this concept), we need to ``merge'' all these components into a unique, global state-machine. This has several implications:
\begin{itemize}
\item  The class attributes of UML state-machines must be merged into a single B state-machine definition. This may require the prefixing of the variable names with the component names to avoid name clashes. The same manipulation has to be done for the operation names (transition labels in UMC) and the other entities that may require duplication.
\item The currently active state of a UML state-machine is represented in B by the current value of an ad-hoc variable \emph{statemachine\_STATE}. There is one such variable for each UML state-machine.
\item Within the B machine structure, all types, constants, and variable definitions and initializations must appear at the beginning of the machine definition. This disrupts the original structure of the system, forcing us to spread the UML state-machine definition into several places in the B machine specification.
\item In UML state-machines, the event pool (a buffer implementing asynchronous communications that contains at each moment the set of signals arrived in a state-machine but not yet dispatched or discarded) is part of the engine support and thus is not explicitly modelled. In B these event-pool components must be explicitly modelled. This is because, contrary to UMC, B is not a tool natively designed for handling UML state-machines. Therefore ``buffer'' variables representing the state-machine event pool are added to the B model. Consequently, the action of sending a signal to another state-machine will be modelled with the insertion of a value to the corresponding variable buffer, and the dispatching of a signal to trigger a transition will be modelled with the extraction of the first element of such a buffer.
\item  Each transition rule definition of the UMC state-machine design is mapped onto an equivalent operation of the B machine.
\end{itemize}

\begin{figure}[ht!]
\centering
\includegraphics[width=0.75 \textwidth]{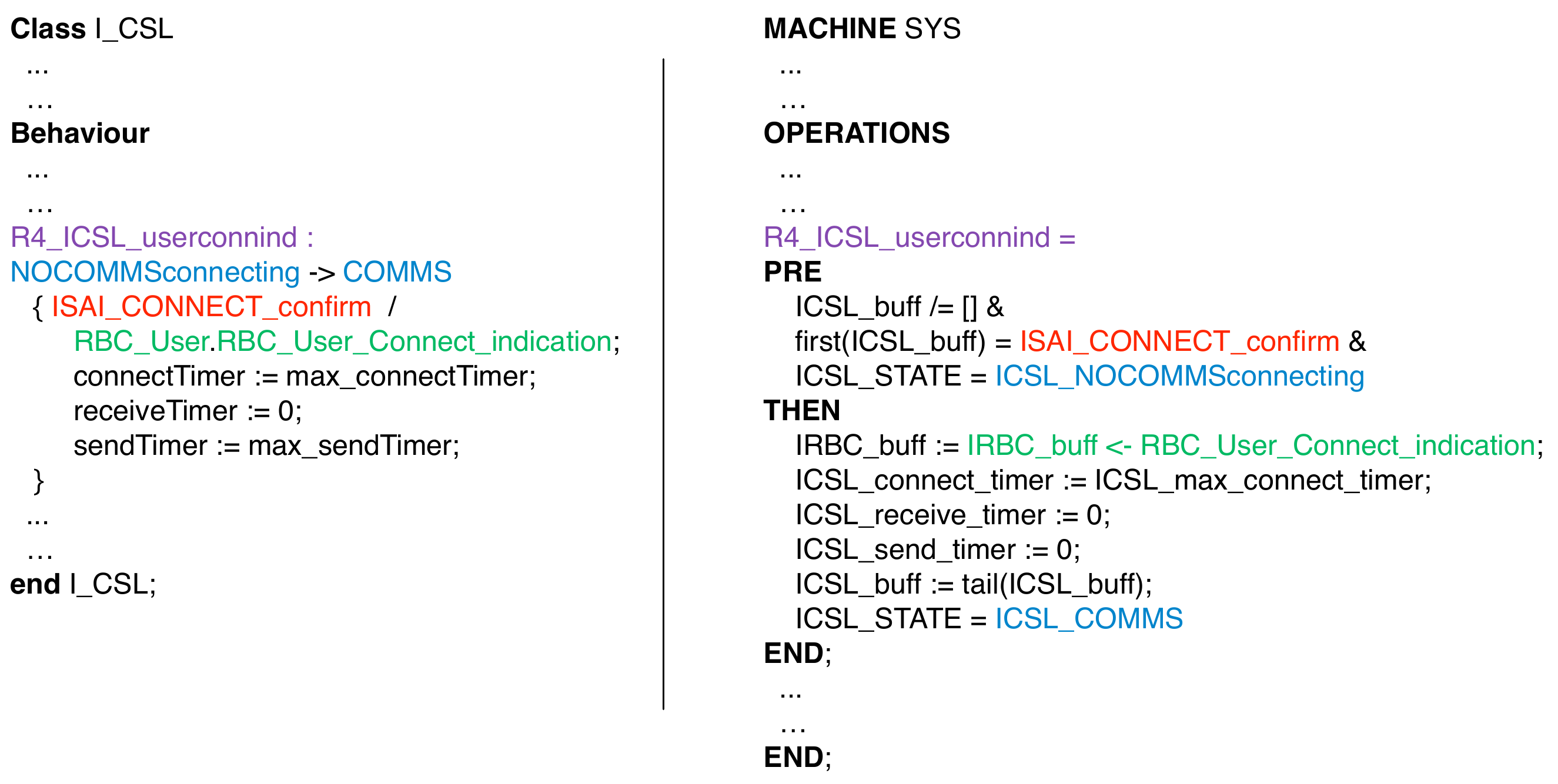}
\caption{Textual encoding of a UMC (left) and ProB (right) state-machine transition}  \label{PROBrule}
\end{figure}

Figure \ref{PROBrule} shows, as an example, the ProB encoding of the UML transition R4 of the initiator CSL component, while the full code  of the state-machine is shown in Appendix C.

\vspace{-10pt}
\subsubsection{CADP/LNT encoding}
LNT is one of the formal notations accepted by the CADP  verification framework.  The notation is a simplified variant of E-LOTOS~\cite{elotos}, of which it preserves the expressiveness but adopts a more user-friendly and regular notations borrowed from imperative and functional programming languages.  A system is described in LNT as a parallel composition of (parametric) processes, which synchronize upon a statically defined set of events. A process can have  private variables that can be manipulated with classical imperative statements.
The global environment is constituted by the data types and functions used by the processes.
An LNT specification is internally translated into the LOTOS~\cite{lotos} algebraic notation and can be analyzed using the CADP toolbox.

\begin{figure}[ht!]
\centering
\includegraphics[width=0.9 \textwidth] {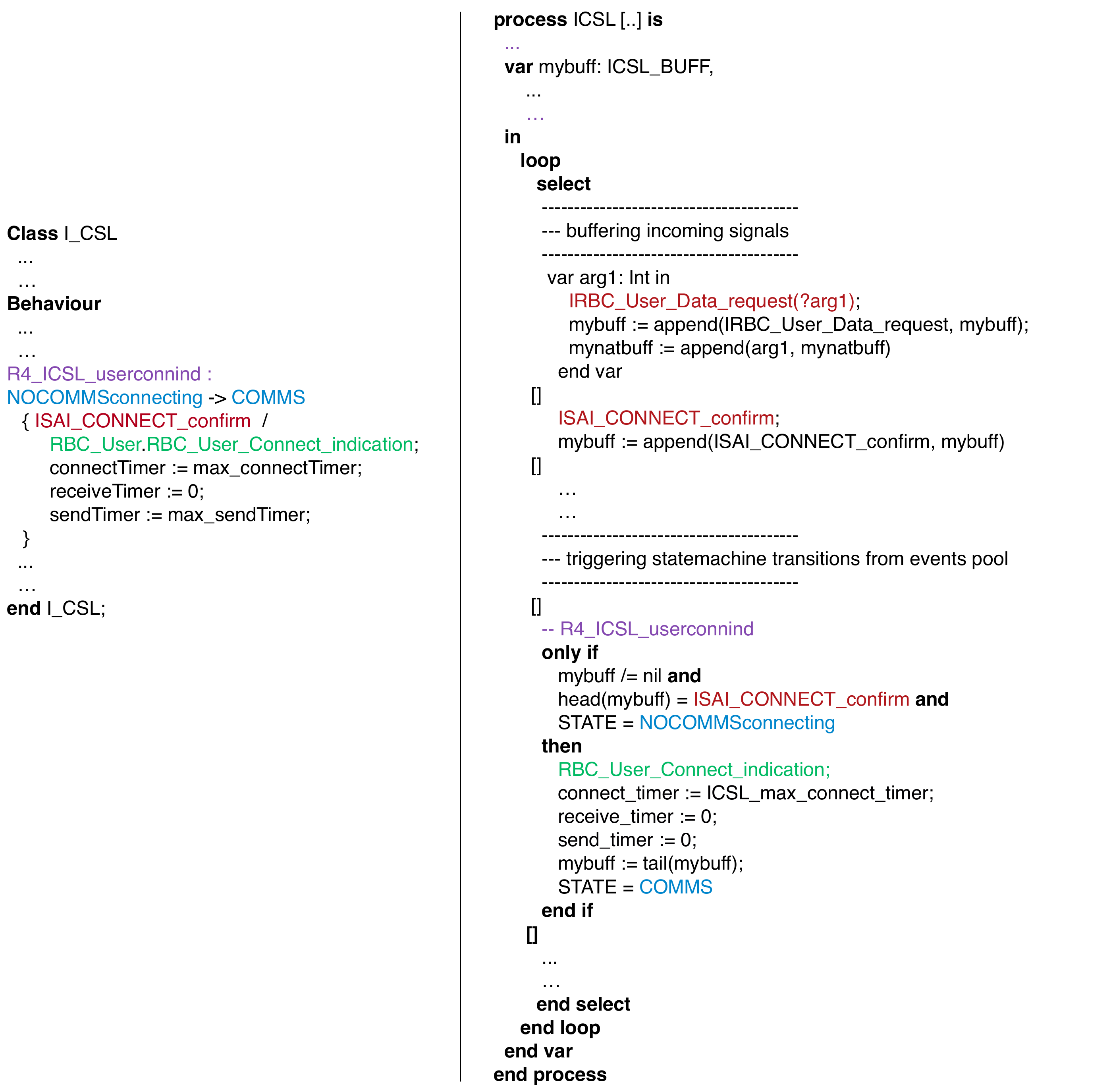}
\caption{The LNT structure corresponding to the initiator CSL state-machine}  \label{LNTprocess}
\end{figure}

In this case, each UMC state-machine is associated with an independent LNT process. All the processes do not share any memory and interact through synchronous actions in the typical style of process algebras. 
Each process handles a local event pool modelled as a FIFO buffer and is \emph{always} enabled to accept synchronizations from other processes willing to push a new event in the queue. Beyond accepting incoming messages, the LNT process can internally evolve, performing internal steps that transform the local status or synchronizing with other processes by sending messages towards other state-machines.
The final system is  obtained by composing in parallel all the processes which synchronize the corresponding actions of sending and receiving a message.
Figure~\ref{LNTprocess} shows the overall structure of a state-machine process corresponding to the initiator CSL, while the full code  of the process is shown in Appendix D.

\vspace {-10pt}
\section{The formal analysis}

The first goal of our analysis has been the proof that all the three generated models are equivalent. This has been done by saving the possible behavior of the models in the form of Labelled Transition System~\footnote{While in the case of CADP and UMC saving a model in the .aut textual LTS is available as a builtin feature of the framework, in the case of ProB this LTS generation has been obtained through a automated transformation of the model state-space originally saved in the ProB ``.statespace'' textual format.}, and by applying comparison tools~\footnote{e.g. mCRL2 ltscompare or CADP bcg_cmp.} to verify that the three ProB and LNT models are strongly equivalent to the UMC models \footnote{UMC can be configured to associate the LTS transition labels with the UMC transition labels, or with the occurring communications actions, or with other observable events.}.
The main goal of the demonstrator, however, is \emph{not} the complete formal verification of a (fragment of a) standard, but the exemplification of the \emph{categories} of costs and benefits that may come to play with the choice of exploiting formal methods for the improvement of system requirements documents.
The focus of our formal analysis is,  therefore, to show with some evidence \emph{how} formal methods may be of help in detecting the design errors potentially introduced while producing the UML executable model, in verifying the high-level properties expected by the full system and by its specific components, and in generating clear and rigorous (graphical) feedback on the specified system to the requirements designers.
The detailed analysis of the costs and benefits, not only qualitative but also as far as possible quantitative, is the object of a separate 4SECURail deliverable \cite{refD2.6}).

From our experience, it has become evident that formal methods can be used in a lightweight (i.e., almost ``push button'') way or in an ``advanced'' way. These two degrees of exploitation of formal methods require a very different level of effort and background. 
A rigorous static analysis of the formal models is probably the simplest example of lightweight use of formal methods. Just loading a system specification in the verification tool may immediately reveal mistakes and anomalies in the code (type violation, non-relevant updates, missing initializations, mismatch of parameters in messages, etc.). 

Other behavioral properties like the absence of deadlocks or examples of reachability of certain states or events can still be verified with just a button-pushing or by writing extremely simple logical properties. Trace examples or counter-examples can be visualised in the form of a UML message sequence diagram \footnote{This is natively possible in the UMC framework and very recently also in the Tck/Tk version of ProB.}. 
Further information can be gathered by monitoring the generation and the statistics on the system state-space (if not too large). The visualization of state-space projections (i.e.,  graphical views of the system state-space once reduced after making observable only some specific detail of the system) can be of great help in understanding and confirming the system behavior without resorting to the encoding of complex temporal logic formulas.

\begin{figure}[ht!]
\centering
\includegraphics[width=0.8 \textwidth]{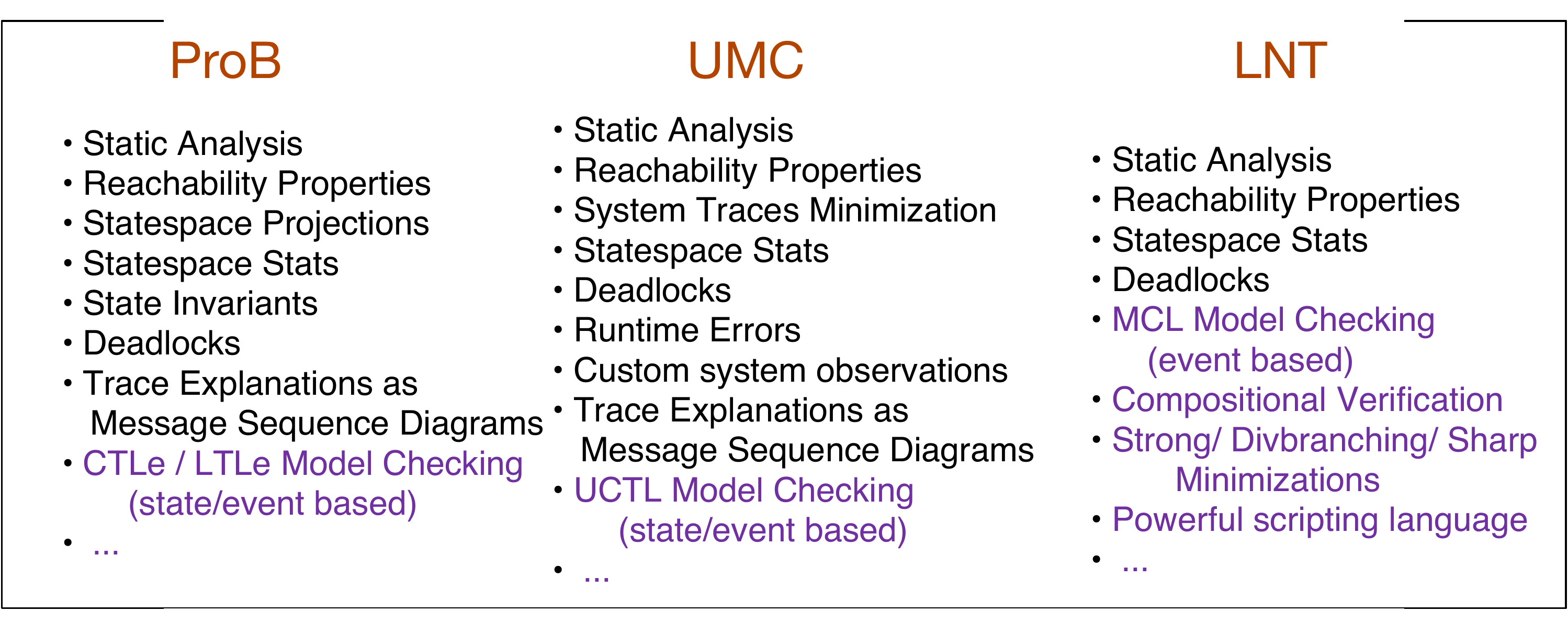}
\vspace {-5pt}
\caption{Table of verification features}  \label{featurestable}
\end{figure}

The analysis of more complex behavioral properties, however, may require the writing of more complex temporal logics formulas. This activity may require a greater background and more advanced knowledge of the verification tools.
Figure~\ref{featurestable} shows a table of \emph{some} of the features provided by our three frameworks. In the table, the features that can be easily exploited without any particularly advanced formal methods and tool knowledge, in an almost ``push button'' way, are those appearing in black.
As it can be seen by observing the mentioned table of features, an advantage of our ``formal methods diversity'' approach is the possibility of exploiting the power offered by the whole set of frameworks,  like state- or event-based model checking, linear- or branching-time model checking, state-space projections, custom system observations, and various state-space minimizations or reductions. 
In our experimentation, the following features have been the most used (more details can be found in Deliverable D2.5 \cite{refD2.5}:
\begin{itemize}
\vspace {-2pt}
\item static analysis in UMC/ProB/LNT
\vspace {-2pt}
\item explanations and animations in UMC/ProB
\vspace {-2pt}
\item weak-complete-divergence-sensitive-trace generation in UMC
\vspace {-2pt}
\item fast state-space generation in UMC
\vspace {-2pt}
\item divbranching minimizations in CADP
\end{itemize}
\vspace {-5pt}
\vspace {-5pt}
When the same feature is available on multiple platforms, also usability aspects play an important role in selecting which one to exploit. E.g. CADP does not allow to observe the evolution of the values of the  process variables during the animation of the behavior or the observation of a counter-example, ProB and UMC have richer visualization system, allowing among the other things to observe a trace in the form of a Sequence Message chart, SVL scripting in CADP makes easier the structuring and documentation of the ongoing verification process.
The downside of this \emph{formal methods diversity} approach is that becoming expert in the use of all these frameworks is likely to require a steep learning curve, with the needed single effort to be multiplied by the number of frameworks and with the risk of not becoming  expert in any of them.

\subsection{Properties and scenarios}

The formal analysis of the system that has been performed during the the project activity is surely not complete, but sufficient to become reasonably confident in the absence of implementation or logical errors. Further tests and verification are still in progress, e.g., from the point of view of compositional verification in the context of the CADP/LNT framework.
Several kinds of architectures can be generated to observe the system properties or the properties of single components.
Figure~\ref{systemcomponents} shows the case of  a ``complete'' architecture, where all the system components are composed together with the needed environment components.  Several flavors of this architecture can be designed, depending on the properties we want to observe, on the limit to the complexity we want to set, and on the kind of behavior of the environment we want to consider.
Once the desired behavior of the environment components is established, they need to be instantiated into an executable scenario with the setting of a list of internal parameters fixing the parametric aspects of the specification.
The simplest architecture we have built is when the two RBC remain ``silent'' (not sending any message) and just receive connection/disconnection indications from the CSL layer. In this architecture, the Euroradio level is imagined to be ``nice'', i.e., introducing at most small delays in the communications,  not losing nor reordering messages, and not autonomously aborting the existing active communication channel.  The set of UML state-machine diagrams describing the components of this architecture is shown in Appendix A.
In this case, the system is simply expected to set up a communication line, keep it alive by exchanging life-signs, and re-establish it in case of failures. Failures can still occur depending on the specific values of the parameters used to instantiate the scenario. In particular, the most important parameters affecting the system behavior in this scenario are:
\vspace {-2pt}
\begin{itemize}
\item The timeout (max\_connectTimer) representing the maximum delay that initiator CSL is allowed to wait before receiving  a reply to a connection request (after which a new connection request can be retried).
\vspace {-2pt}
\item The timeout (max\_initTimer) representing the maximum delay that each SAI is allowed to wait for the successful initialization of a new communication line before aborting the creation process.
\vspace {-2pt}
\item The timeout (max\_sendTimer) triggering the periodic sending by the CSL of a new life-sign to keep the communication line alive.
\vspace {-2pt}
\item The timeout (max\_receiveTimer) representing the maximum delay  a CSL is allowed to wait before receiving a life-sign or a rbc message from the other side (whose expiration causes the abort of the re-establishment of the communication line).
\end{itemize}
Other important system parameters, like
\begin{itemize}
\item The limit (N) of consecutive loss of messages (detected by the observation of sequence numbers) acceptable by the SAI components, before aborting the safe connection line.
\vspace {-2pt}
\item The maximum traveling delay (K) acceptable for incoming messages whose violation forces the discarding of the message.
\end{itemize}
do not play a relevant role in this scenario.

In this case, we can observe that if the connection, initialization, and receive timer are sufficiently large\footnote{time unit are measured as multiples of the basic system execution cycle.} (e.g., max\_connectTimer = max\_initTimer = 20,  max\_receiveTimer = 15, max\_sendTimer = 5) the system successfully establishes an initial communication line without ever losing it.
If we instead reduce the max\_receiveTimer parameter to 8, communication failures and communications line restarts begin to appear
(and the system state-space grows from 19,788,895 to  74,713,472 states).

An extension of the previous architecture is where the RBCs are allowed to send slots of ``nmax'' messages.  In this case, we can observe how messages, if they arrive, are delivered to the target RBC without reordering, duplications, and within a maximum delay. In this case, the system state-space grows to 65,386,049 and to 84,883,327 states when slots of 1 or 2 messages are sent by just the RBC on the initiator side.
Beyond the complete architectures described above,  other kinds of architectures have been set up.
For example, an ``ICSL testing'' architecture, where the Initiator CSL component is stimulated with an abstract model of the SAI and RBC components, and an ``Initiator-side testing'' architecture, where the whole ER layer and  CSL/SAI/RBC on the ``called side'' of the system are abstracted by environment components. In this latter case, we can observe how the messages received from the RBC environment, in the absence of disconnections, are always delivered to the EuroRadio level without losses, duplications, reordering, and within a limited delay.

Further examples of the verifications that have been done on the models can be found on \cite{refD2.5}.


\section{Related works}
The experimentation of formal methods diversity for the analysis of the same specification has already been described by one of the authors in \cite{refST7T7,refMARSten}. In that case, the focus was on a much simpler case study that did not have the complexity of a parametric signaling standard.  
As a collateral activity of the project, the same fragment of UNISIG SUBSET 98 has been modelled and verified with UPPAAL by Basile et al. in \cite{refFMICSbasile}. 
The current translation of UML state-machine diagrams into ProB has been initially experimented in \cite{refD43}, but other approaches are possible; the UML-like UML-B notation \cite{refUMLB,refSYSMLB} has been proposed as a  bridge between an Eclipse-based model framework (Rodin) and the Event-B modeling notation; the suggested approach seems, however, to be tailored to the verification and refinement of single state-machines and not to the analysis of the overall behavior of a set of interacting state-machines.
Many other formal notations have been the target of translations from SysML design. Another work very similar to our from the point of view of the goal is the one described by Bouwman et al. \cite{refBasPoint}. Also in that case the goal was aimed at the analysis of a signaling standard under development rather than the verification of a specific system. The target notation and framework is, in that case, mCRL2.

\section{Conclusions}

Formal analysis of a still fluid, parametric, and environment-depending requirements specification (i.e., requirements elicitation and validation) is a very different kind of activity than  verifying that a given implementation is correct with respect to a specific, stable, and rigorous specification.  
The possibility to exploit the analysis features offered by more than one verification framework can be of help in approaching this activity. 
Moreover, the design of several different scenarios can be necessary to observe the system behavior under various assumptions. From this point of view, the possibility to \emph{automatically} generate the formal models to be analyzed  from some executable, widely known, standard, tool-independent notation is a crucial point  to make the analysis process accessible also from to people with the relevant railway-signaling knowledge.
The formal methods diversity approach experienced in the project has shown how a lightweight use of formal verification frameworks can already, with a small effort, produce important feedback on the quality of the design. A deeper and more advanced exploitation of all the available features, however, remains a difficult and daunting task, especially when the system complexity and size grow to a level requiring ad hoc mitigation approaches.
The activity shown with our experimentation can be continued and improved in several directions. The executable UML subset used in the project can be greatly extended, still preserving its clear and rigorous semantics and its possibility of automatic translation into several formal notations. 
Also, the set of target formal notations (currently limited to UMC, ProB, and LNT) can be extended with a likely small effort to further frameworks like mCRL2, Spin, nuXmv, just to mention some. The detailed description of the project results, the initial executable UML designs, their formal encoding, the source code of the translators, are all publicly available \cite{ref4SECdeli,refZENmodels,refZENcode}.

\vspace{10pt}
{\small
\textbf{Acknowledgements}

This work has been partially funded by the 4SECURail project.
The 4SECURail project received funding from the Shift2Rail Joint Undertaking under the European Union's Horizon 2020 research and innovation programme under grant agreement No 881775 in the context of the open call S2R-OC-IP2-01-2019, part of the ``Annual Work Plan and Budget 2019", of the programme H2020-S2RJU-2019. 
The content of this paper reflects only the authors' view and 
the Shift2Rail Joint Undertaking is not responsible for any use that may be made of the included information.
We are grateful to the colleagues of the Work Stream 1 of project 4SECURail, and in particular to Alessandro Fantechi, Stefania Gnesi, Davide Basile, Alessio Ferrari, Maurice ter Beek, Andrea Piattino, Laura Masullo and Daniele Trentini for the comments and suggestions during the project.}
\pagebreak
\nocite{*}
\bibliographystyle{eptcs}
\bibliography{mars}
\pagebreak
{\Large \textbf{Appendix A:  UML diagrams for all the  CSL and SAI components}}

\begin{figure}[!htb]
\centering
\includegraphics[width=0.85 \textwidth]{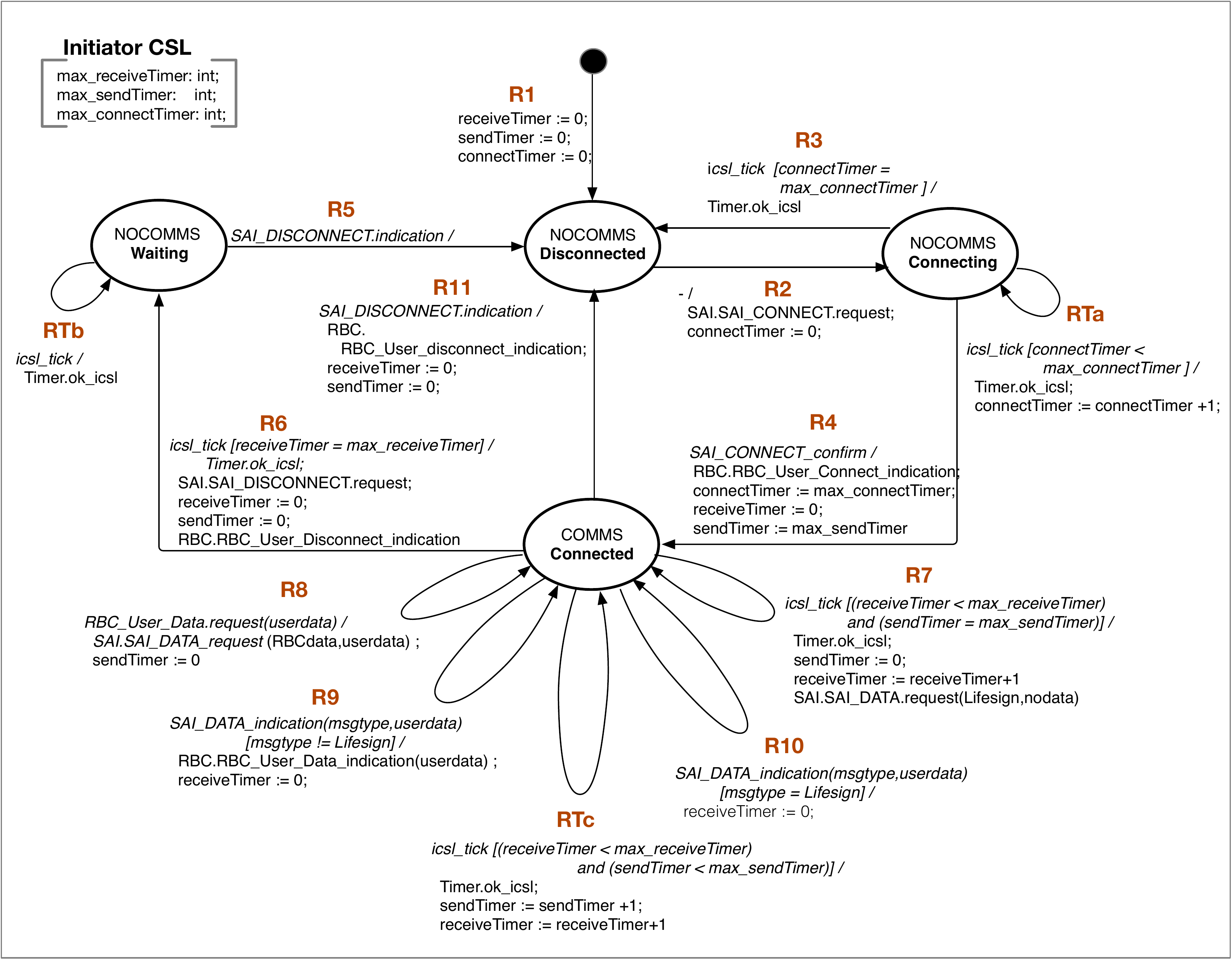}
\caption{The Initiator CSL state-machine}  \label{icsl}
\end{figure}

\begin{figure}[!htb]
\centering
\includegraphics[width=0.75 \textwidth]{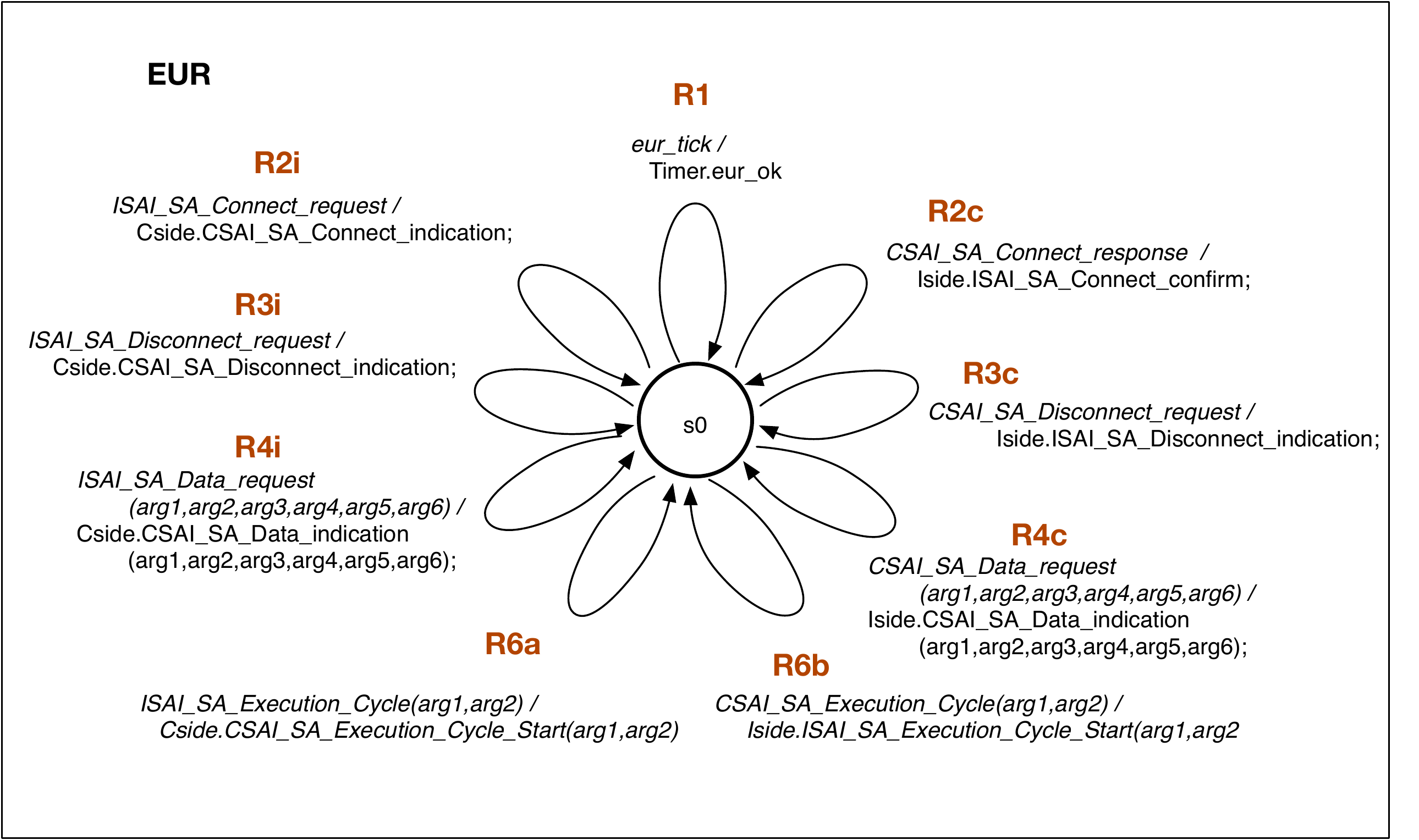}
\caption{The \emph{nice} Euroradio component of the Full scenario}  
\label{euroradio}
\end{figure}

\begin{figure}[!htb]
\centering
\includegraphics[width=0.9 \textwidth]{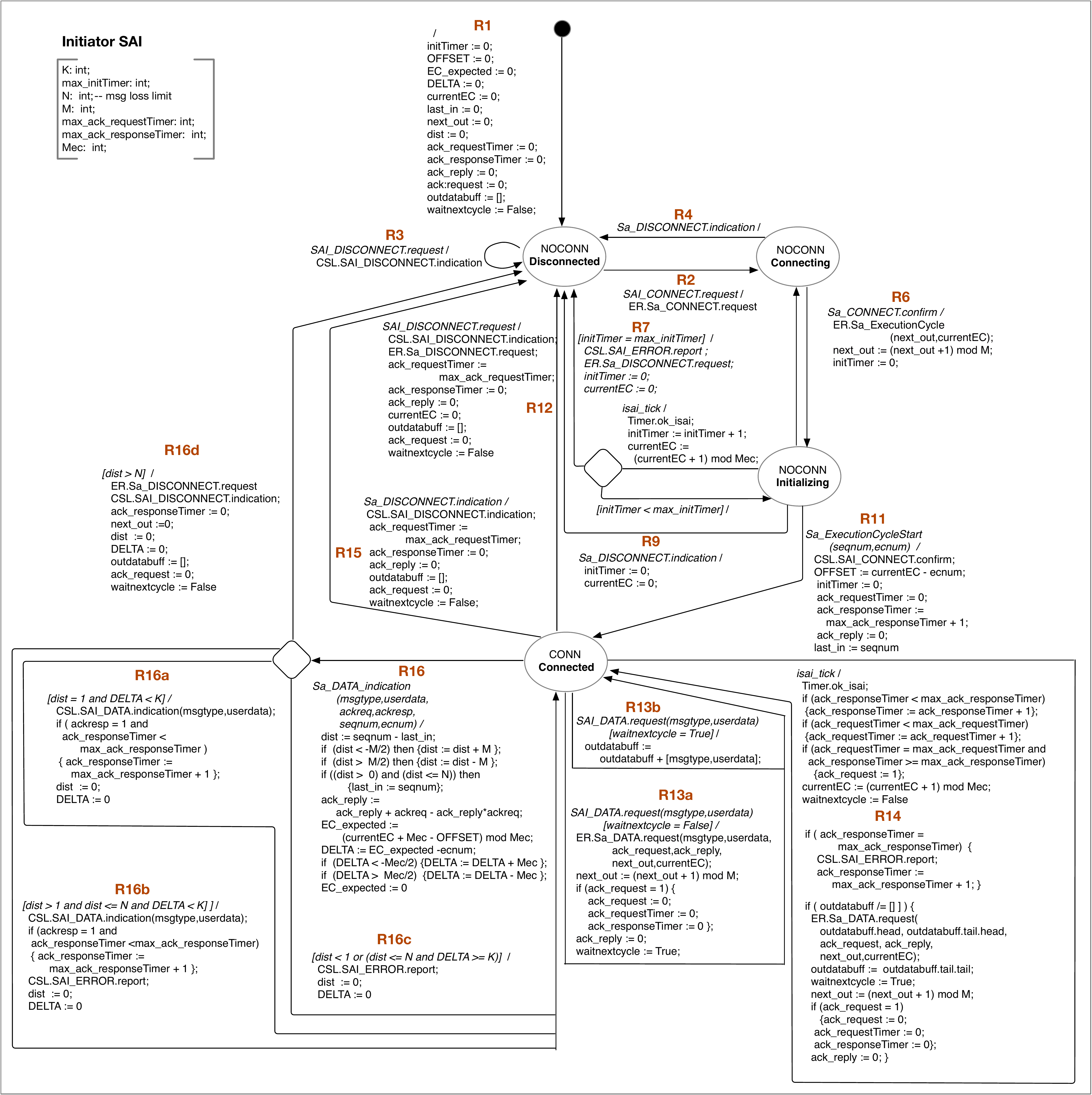}
\caption{The Initiator SAI state-machine}  \label{isai}
\end{figure}

\begin{figure}[!htb]
\centering
\includegraphics[width=0.5 \textwidth]{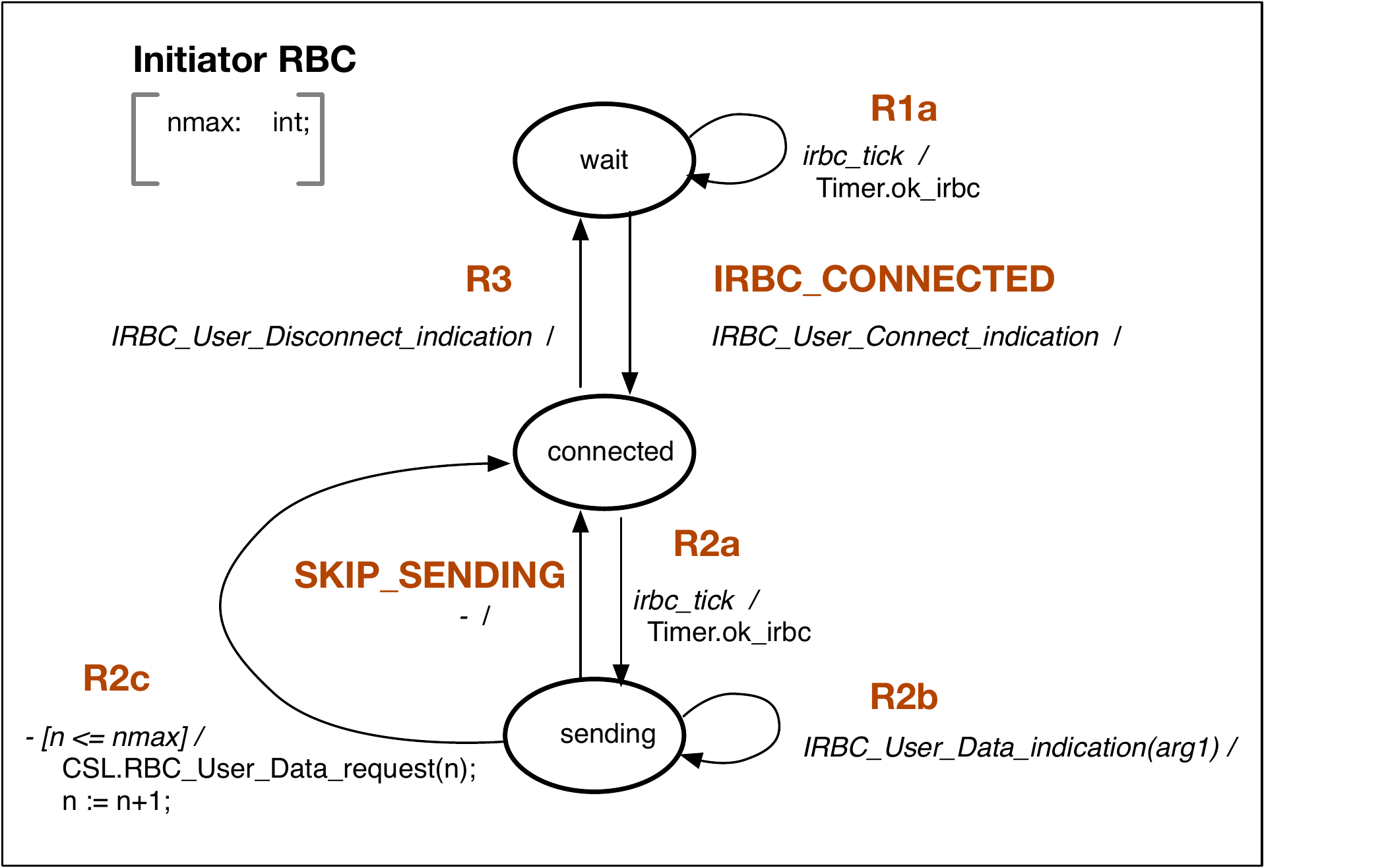}
\caption{The Initiator-side RBC state-machine}  
\label{irbc}
\end{figure}

\begin{figure}[!htb]
\centering
\includegraphics[width=1 \textwidth]{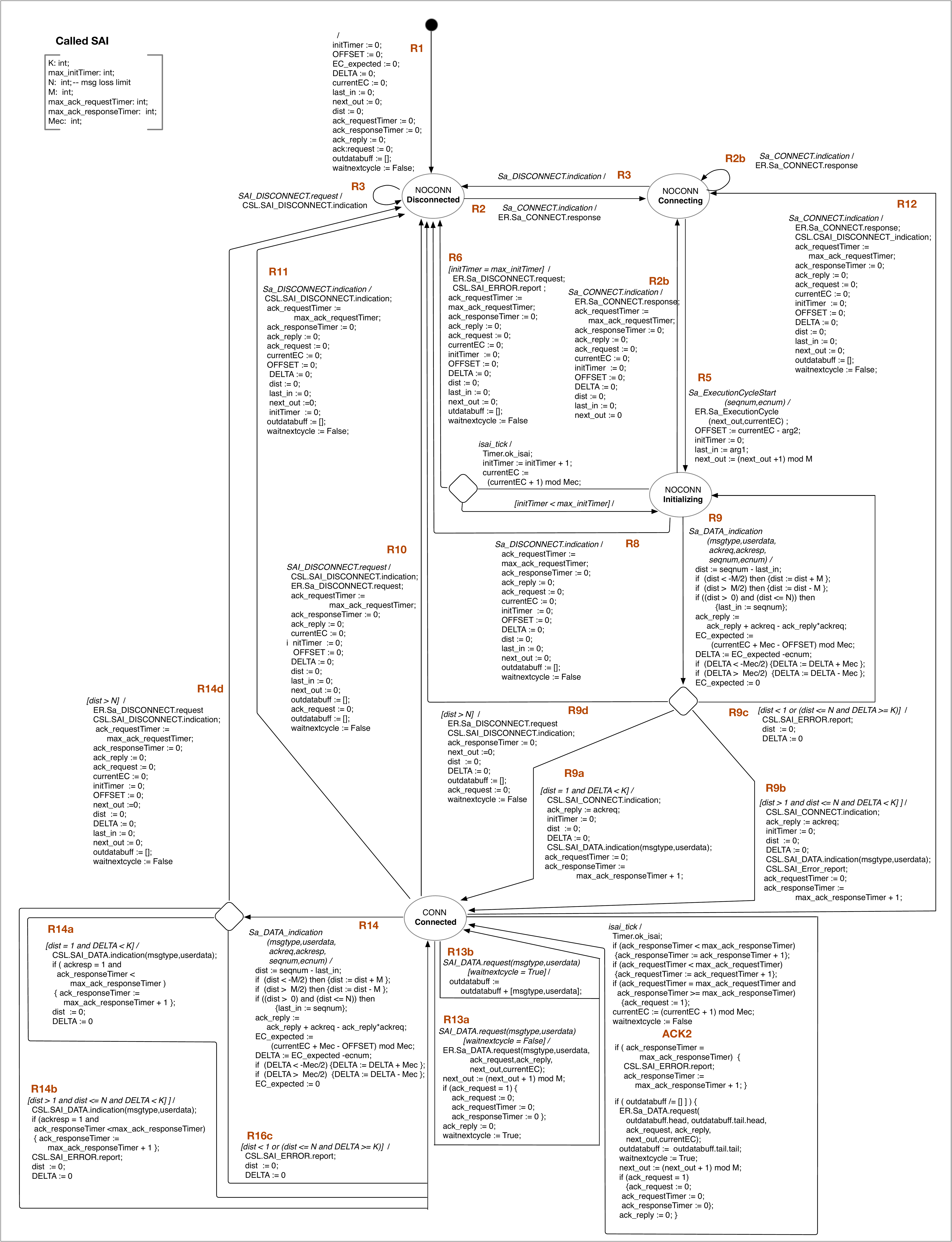}
\caption{The Called SAI state-machine}  \label{csai}
\end{figure}

\begin{figure}[!htb]
\centering
\includegraphics[width=0.7 \textwidth]{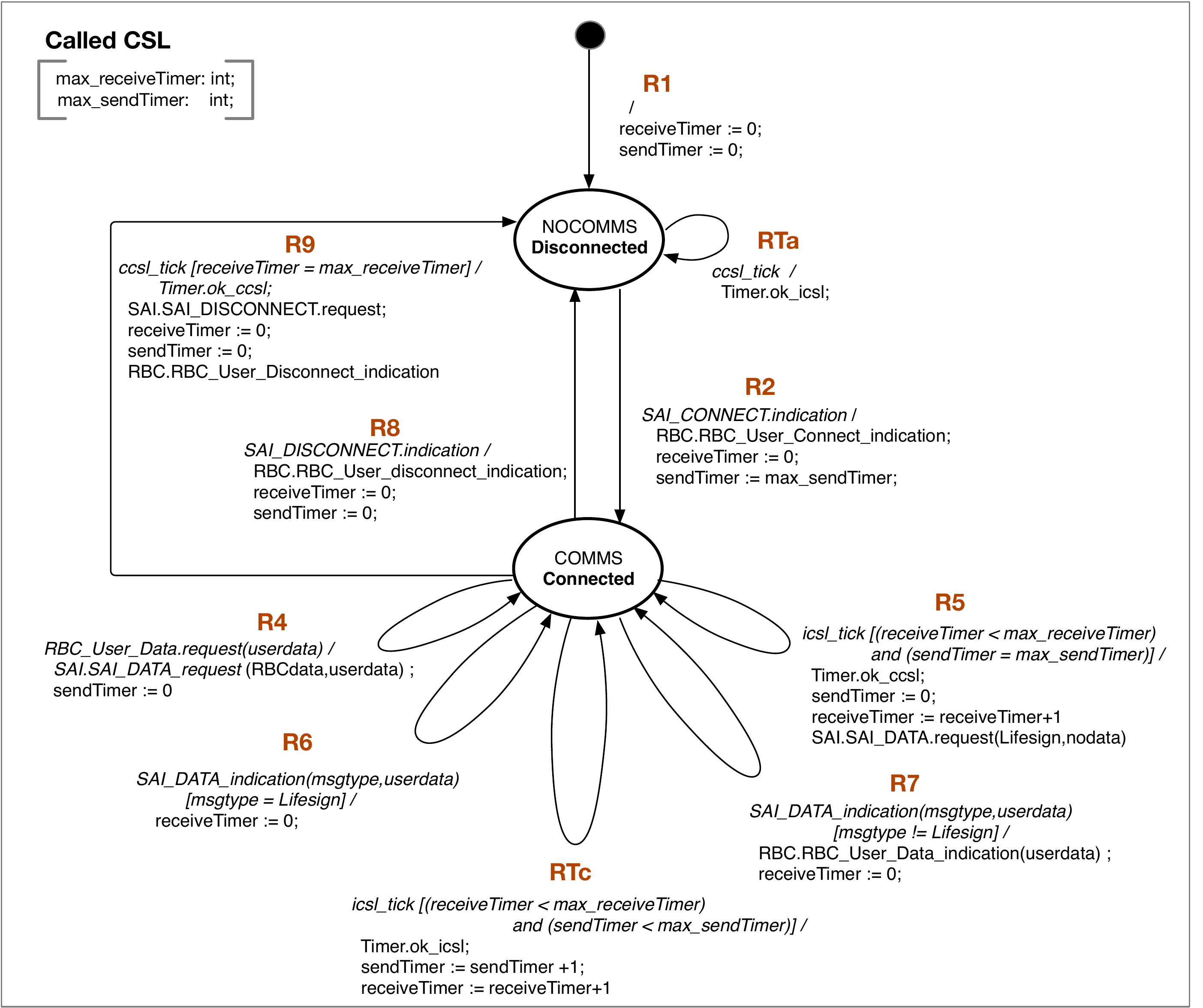}
\caption{The Called CSL state-machine}  \label{ccsl}
\end{figure}

\begin{figure}[!htb]
\centering
\includegraphics[width=0.55 \textwidth]{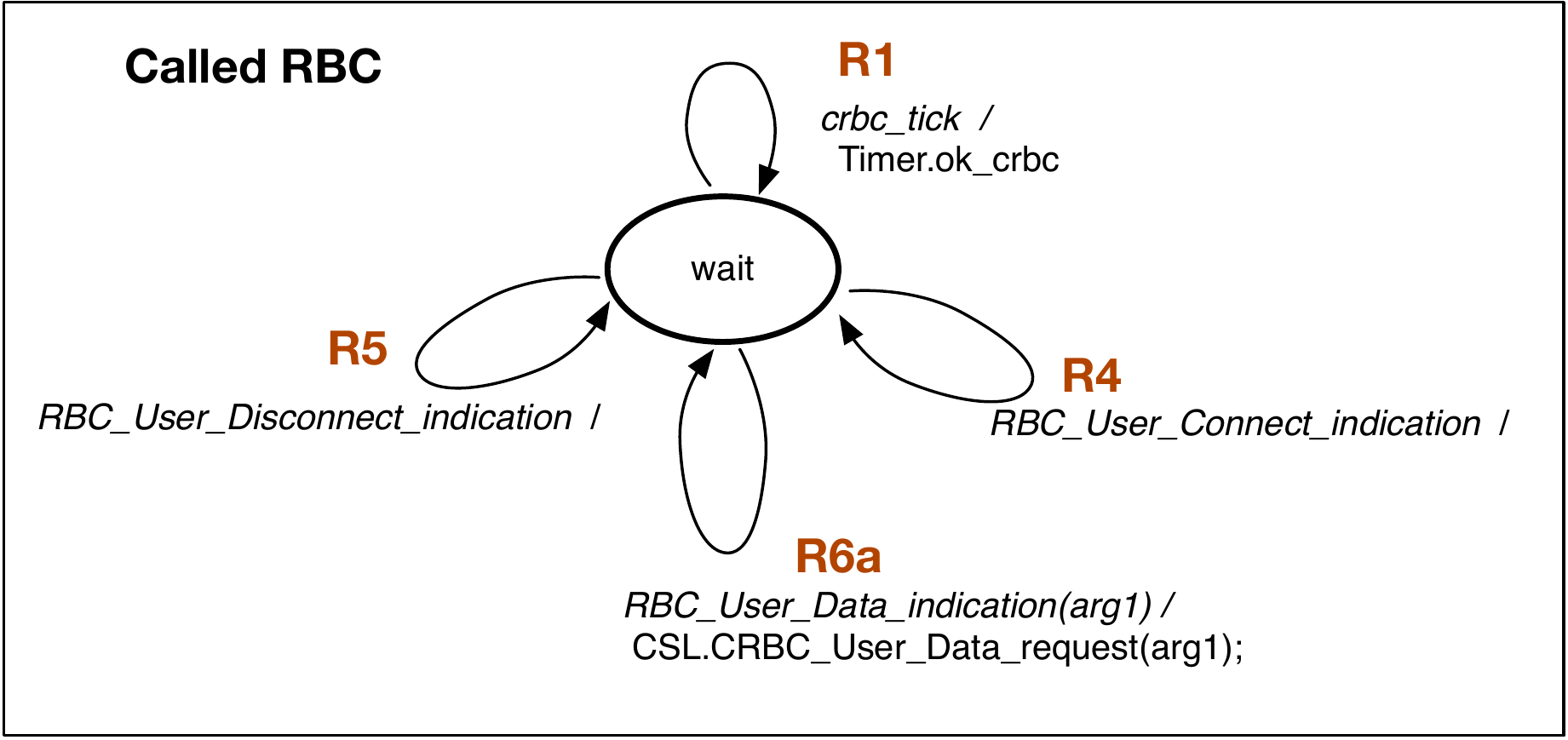}
\caption{The Called-side RBC state-machine}  \label{crbc}
\end{figure}

\begin{figure}[!htb]
\centering
\includegraphics[width=0.4 \textwidth]{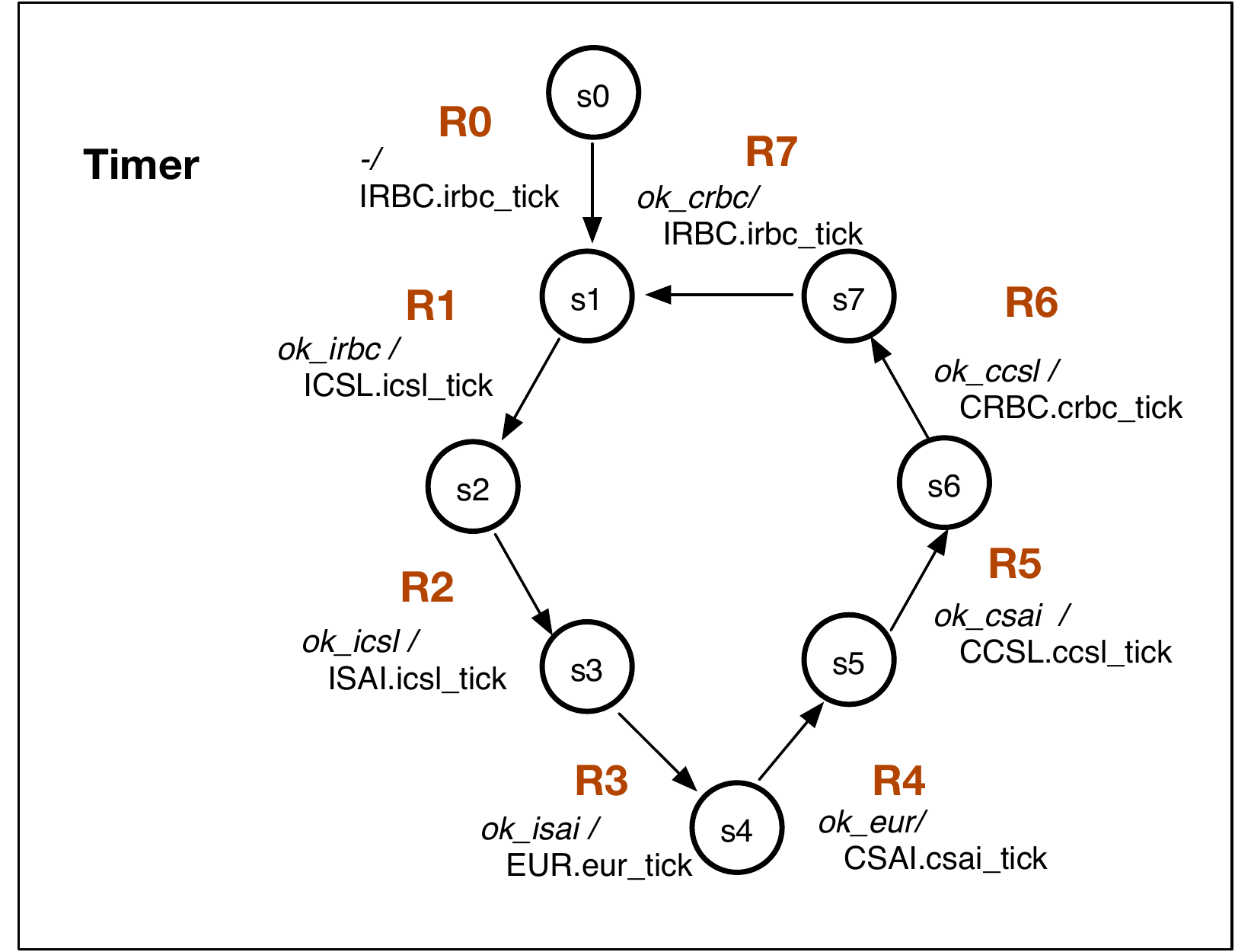}
\caption{The Timer component of the Full scenario}  \label{timer}
\end{figure}

\clearpage 
{\Large \textbf{Appendix B:  UMC encoding of the initiator CSL class  }}
\begin{figure}[!htb]
\centering
\includegraphics[width=0.85 \textwidth]{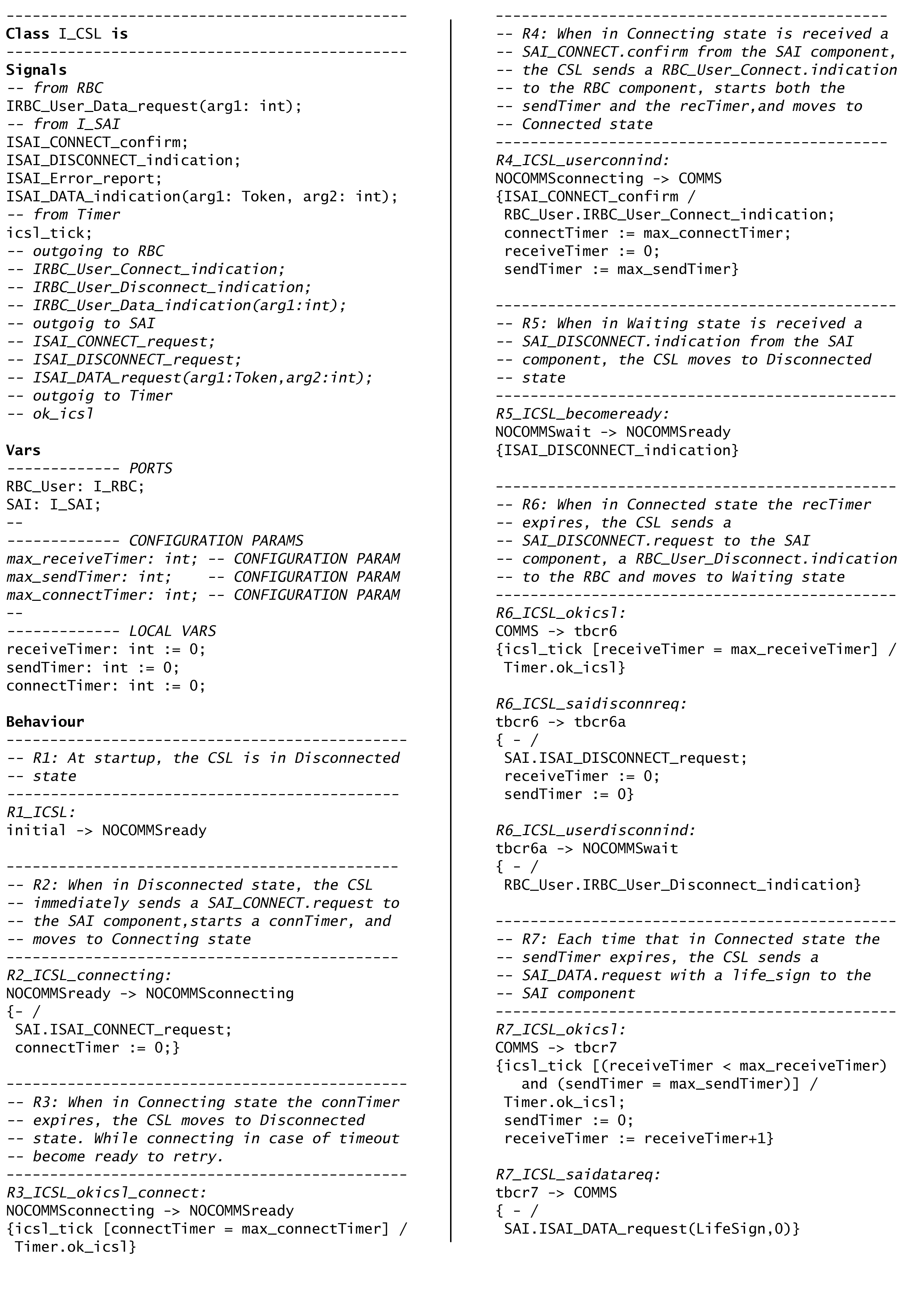}
\end{figure}

\begin{figure}[!htb]
\centering
\includegraphics[width=0.85 \textwidth]{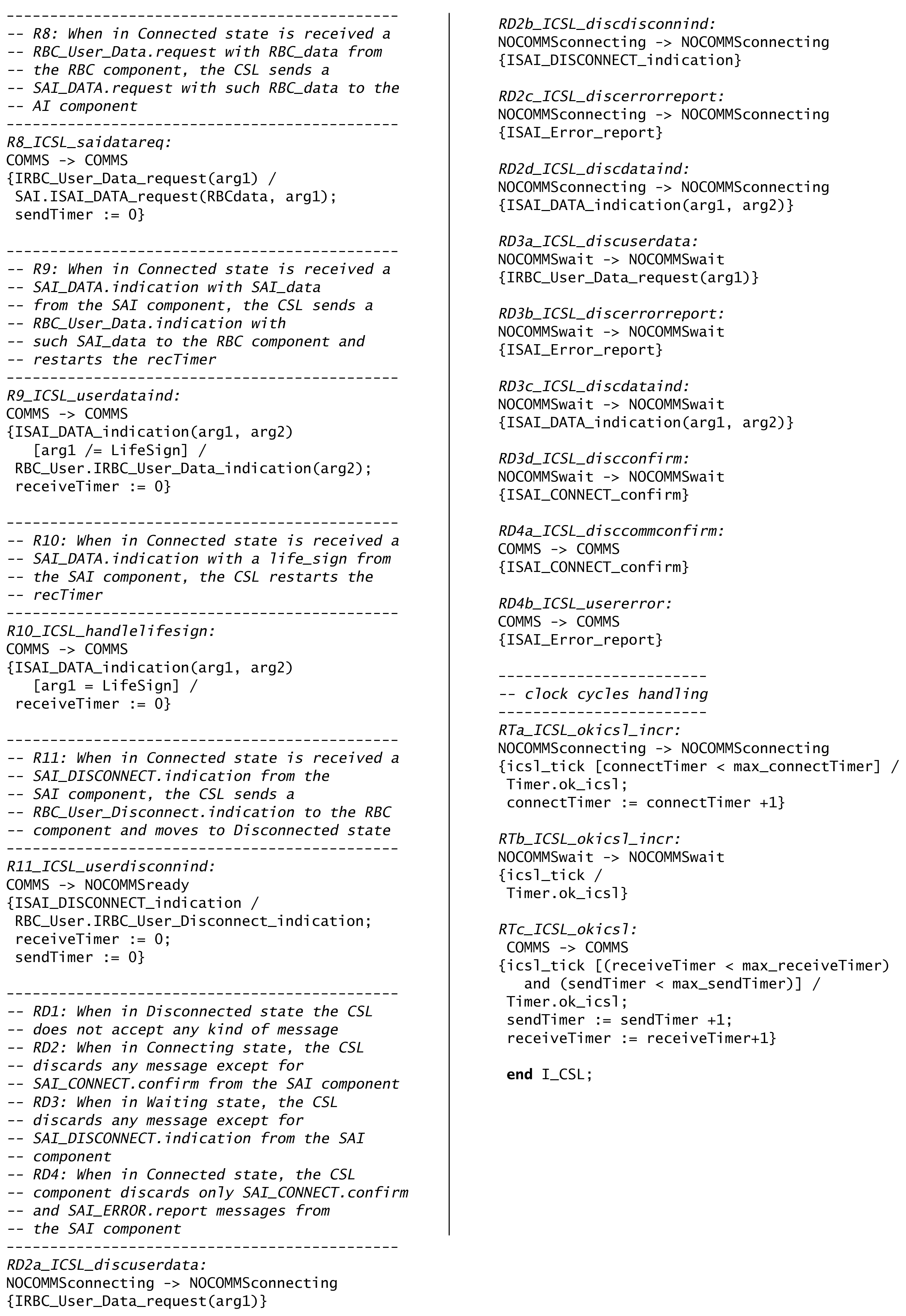}
\end{figure}

\clearpage 
{\Large \textbf{Appendix C:  ProB encoding of the initiator CSL class }}
\begin{figure}[!htb]
\centering
\includegraphics[width=0.85 \textwidth]{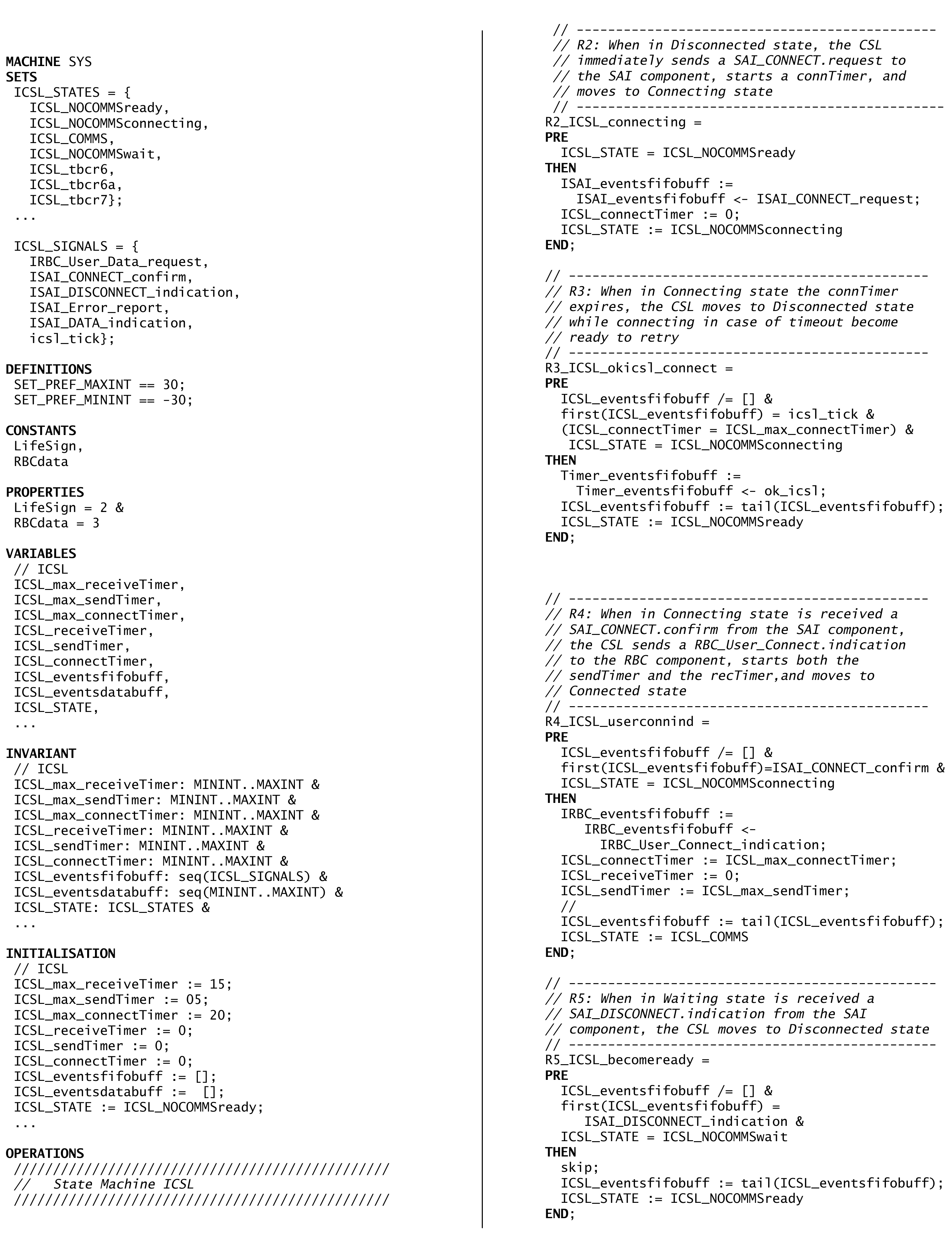}
\end{figure}

\begin{figure}[!htb]
\centering
\includegraphics[width=0.85 \textwidth]{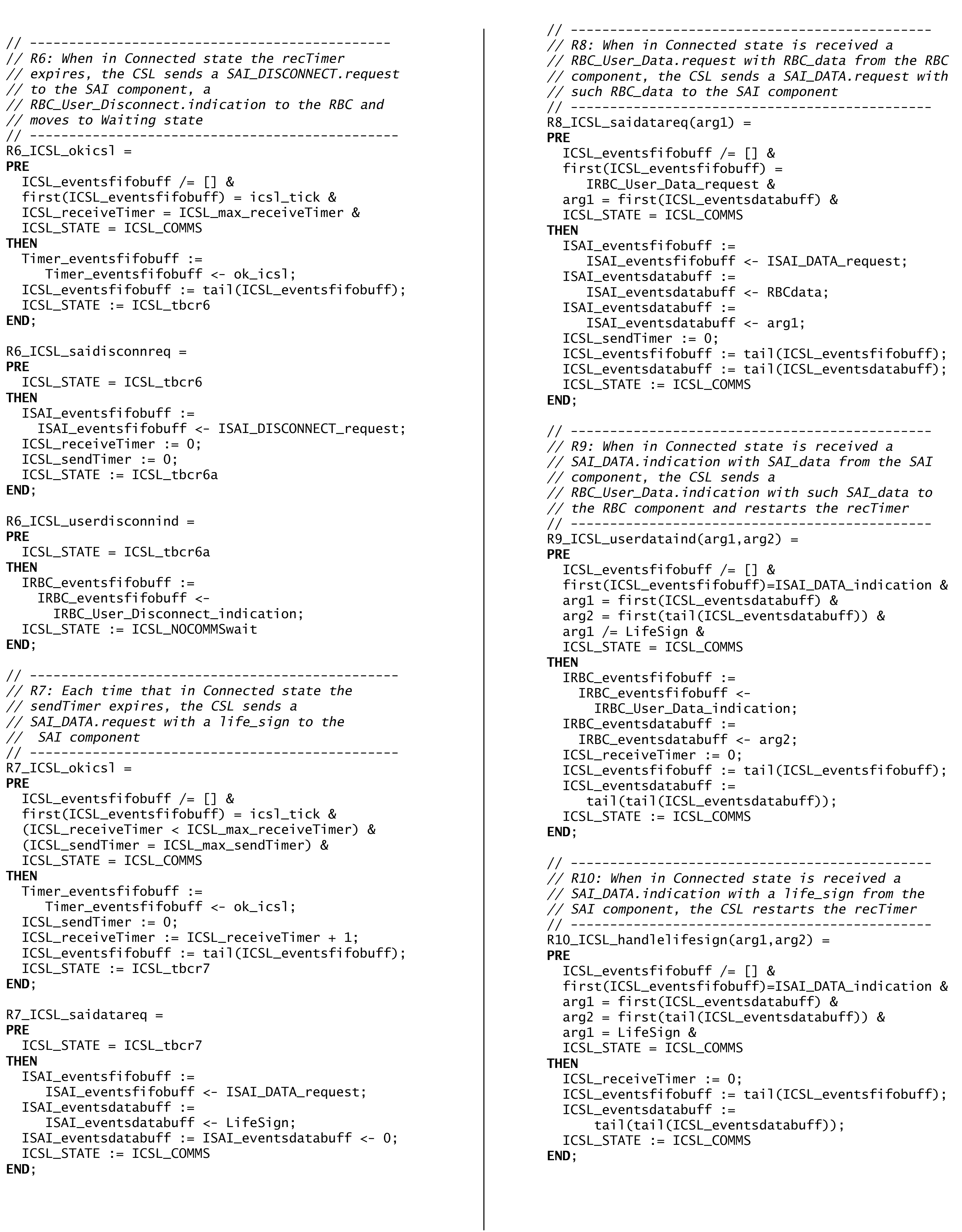}
\end{figure}

\begin{figure}[!htb]
\centering
\includegraphics[width=0.85 \textwidth]{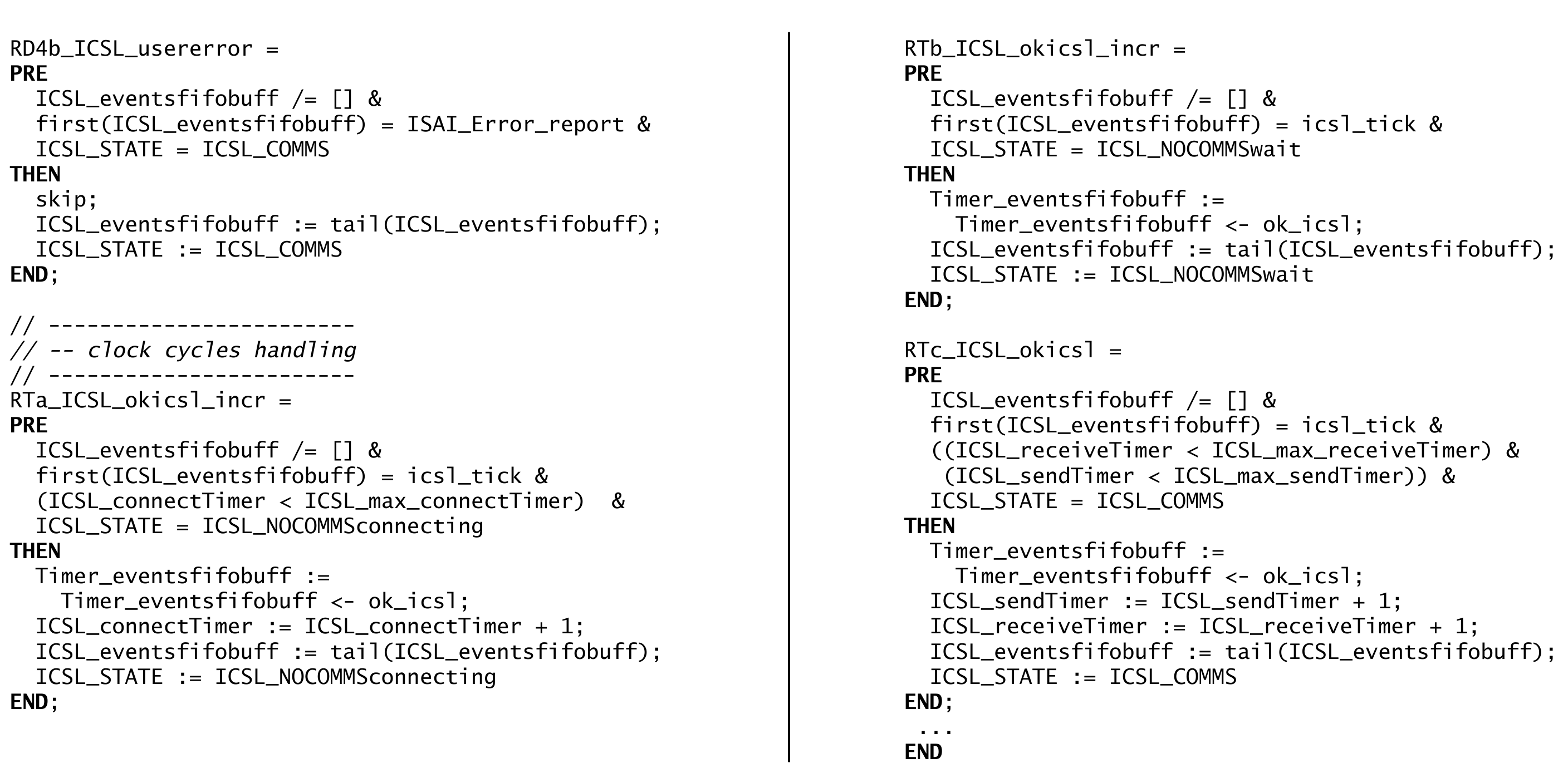}
\end{figure}

\clearpage 
{\Large \textbf{Appendix D:  LNT encoding of the initiator CSL class }}

\begin{figure}[!htb]
\centering
\includegraphics[width=0.85 \textwidth]{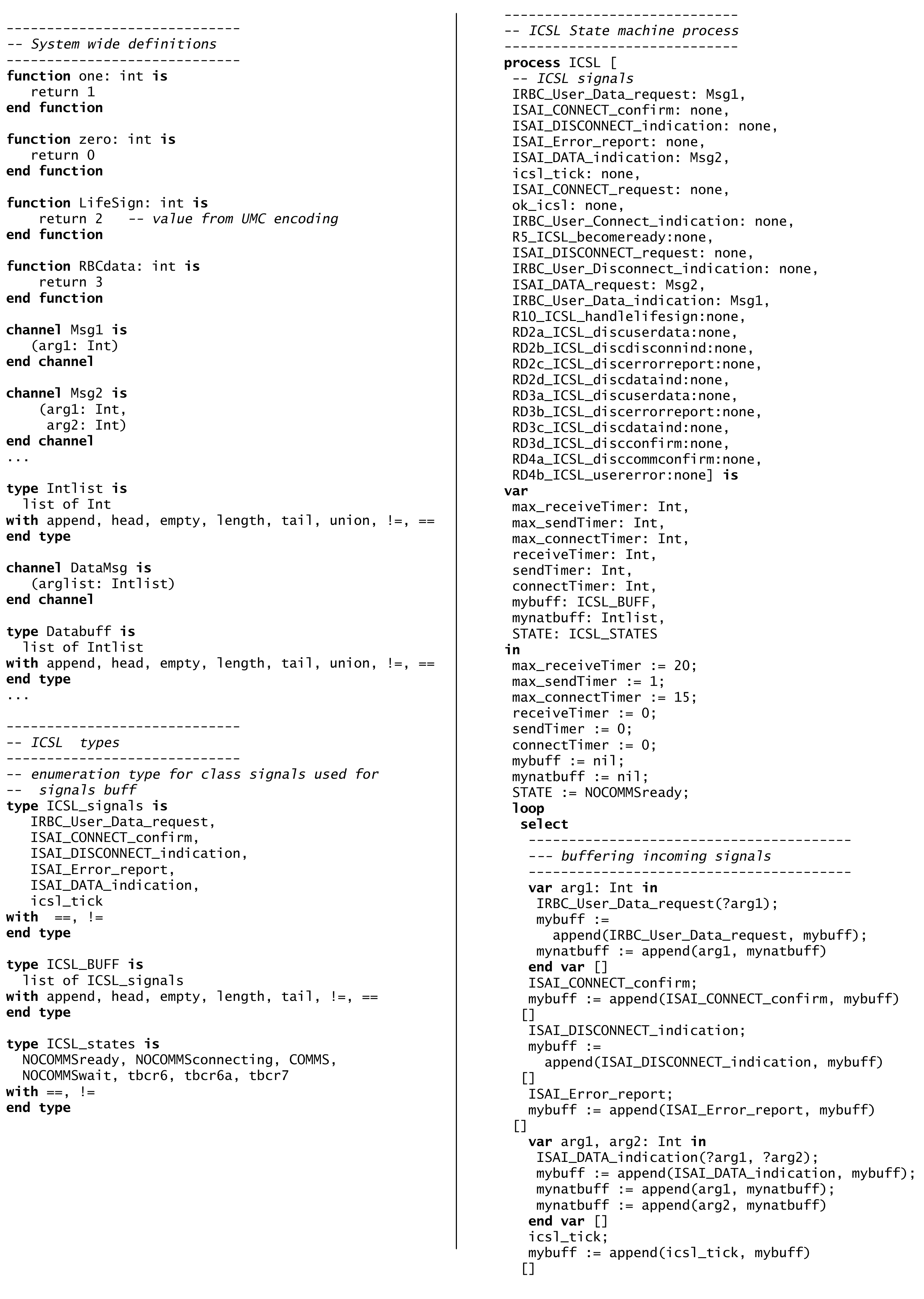}
\end{figure}

\begin{figure}[!htb]
\centering
\includegraphics[width=0.85 \textwidth]{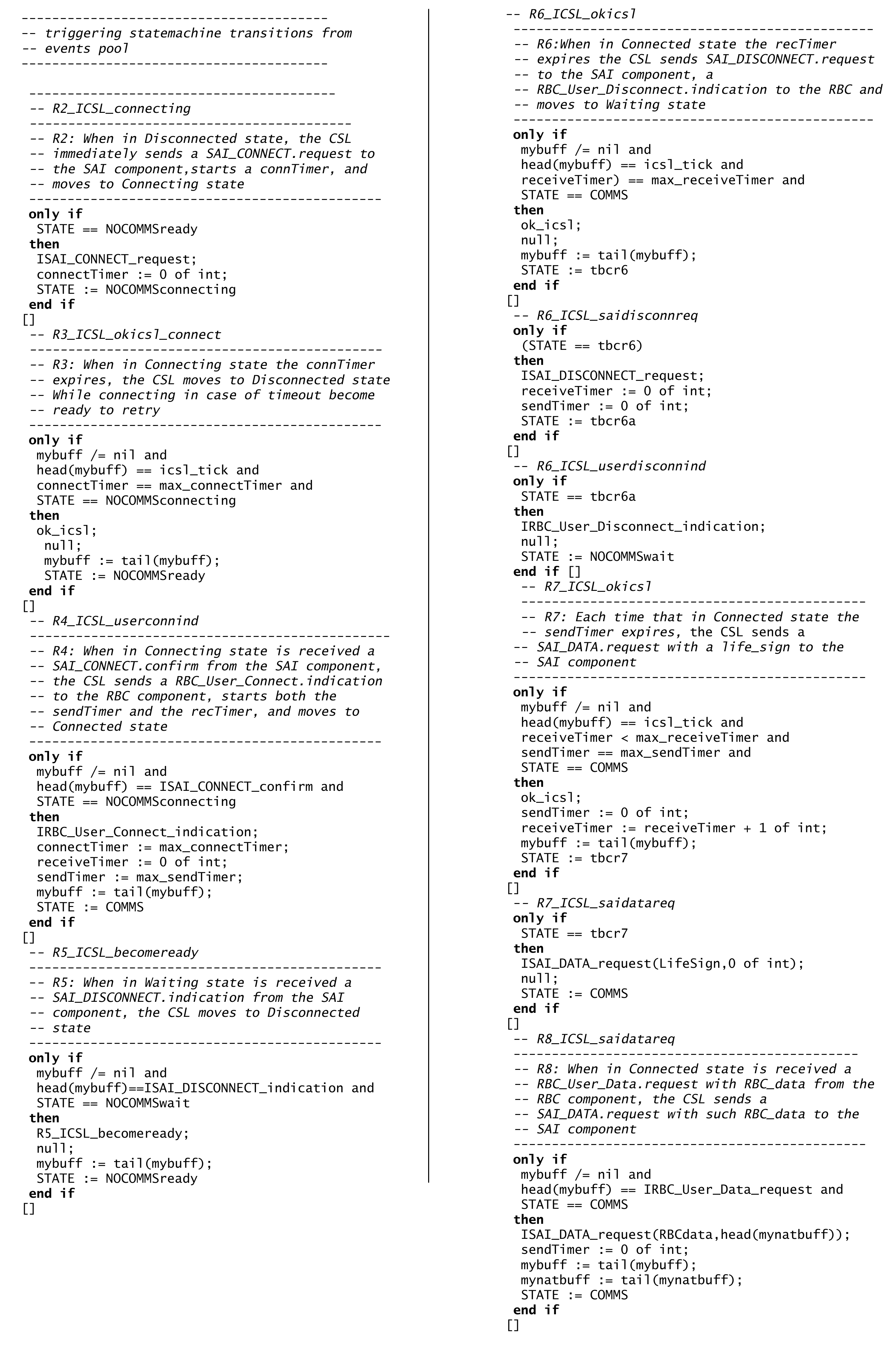}
\end{figure}

\begin{figure}[!htb]
\centering
\includegraphics[width=0.85 \textwidth]{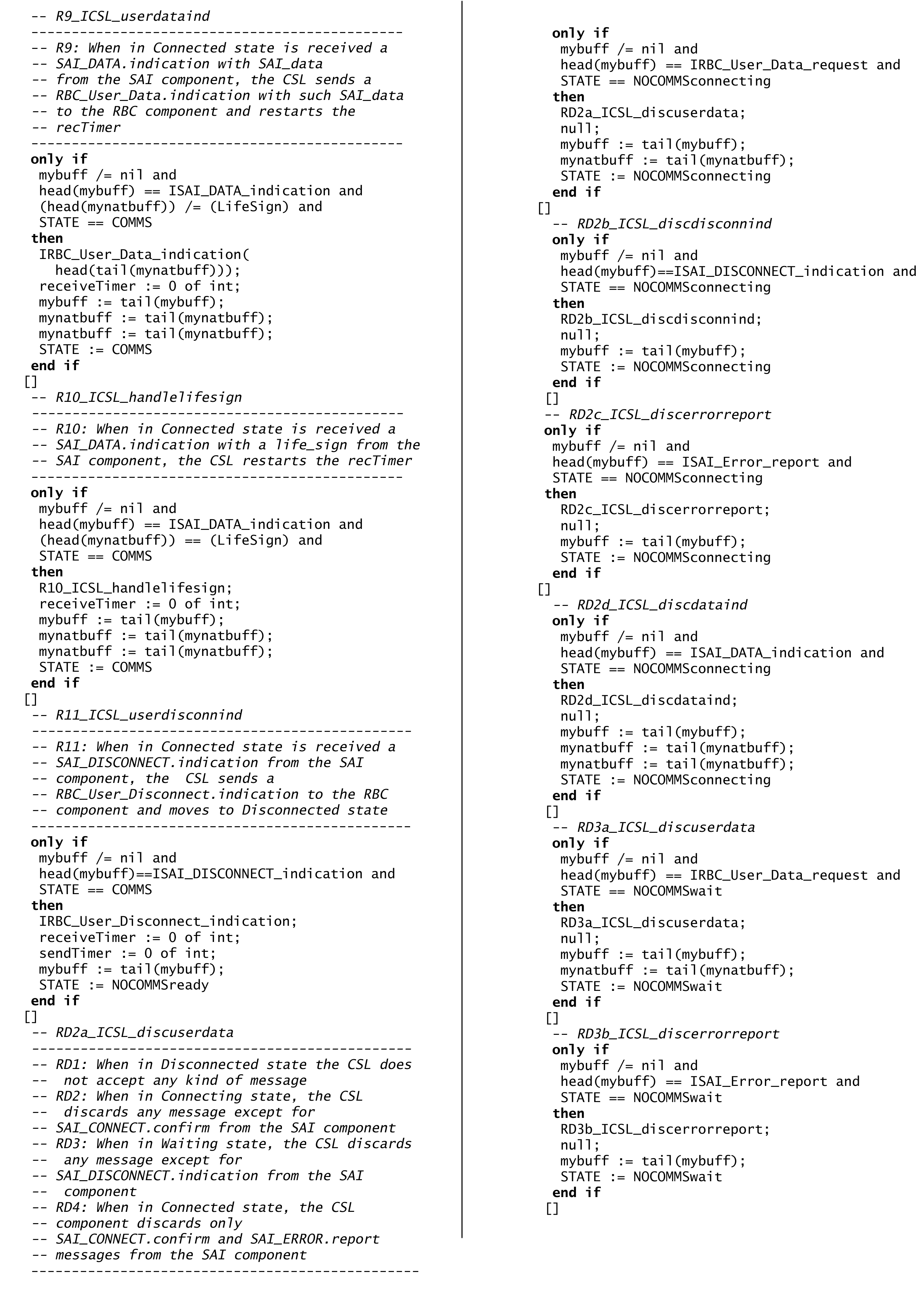}
\end{figure}

\begin{figure}[!htb]
\centering
\includegraphics[width=0.85 \textwidth]{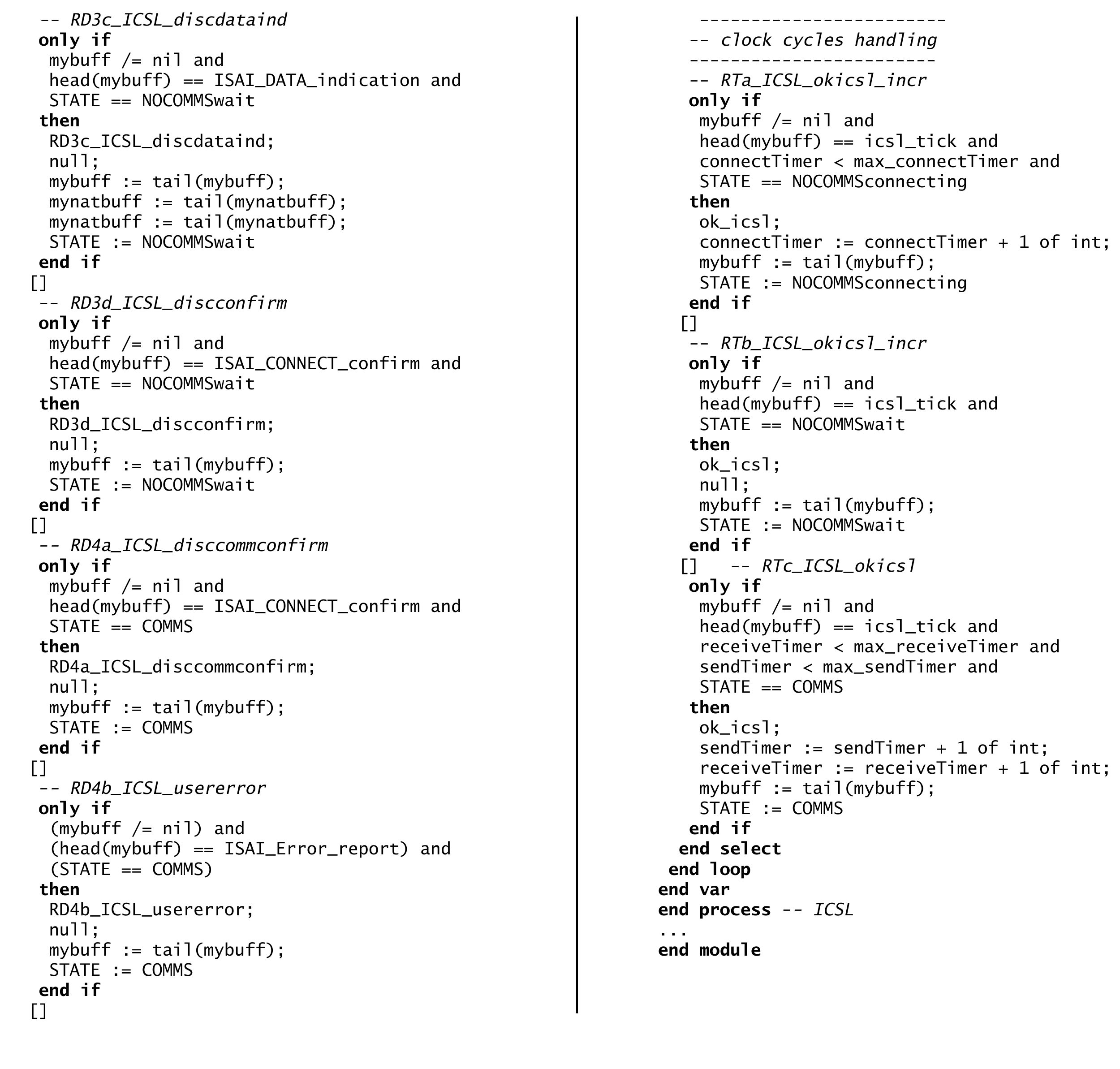}
\end{figure}

\end{document}